# Analysis of a Cyclotron Based 400 MeV/u Driver System For a Radioactive Beam Facility

By

F. Marti, R.C. York, H. Blosser, M.M. Gordon,
D. Gorelov, T. Grimm, D. Johnson, P. Miller,
E. Pozdeyev, J. Vincent, X. Wu, A. Zeller

# 1 Introduction

The creation of intense radioactive beams requires intense and energetic primary beams. A task force analysis of this subject recommended an acceleration system capable of 400 MeV/u uranium at 1 particle µA (6.2 x $10^{12}$ particles/s) as an appropriate driver for such a facility. Further, the driver system should be capable of accelerating lighter ions at higher intensity such that a constant final beam power (≈100 kW) is maintained. This document is a more detailed follow on to a previous analysis of a system incorporating a cyclotron.[1] The driver pre-acceleration system proposed here consists of an ion source, radio frequency quadrupole (RFQ), and linac chain capable of producing a final energy of 30 MeV/u and a charge (Q) to mass (A) of Q/A ≈1/3. This acceleration system would be followed by a Separated Sector Cyclotron with a final output energy of 400 MeV/u. This system provides a more cost-effective solution in terms of initial capital investment as well as operation compared to a fully linac system with the same primary beam output parameters.

# 2 30 MeV/u Pre-Accelerator

## 2.1 General Description

The maximum energy for the pre-accelerator system of 30 MeV/ u is determined by the need to meet the requirement of energy variability for the light ions, particularly protons. The rf frequency of the Separated Sector Cyclotron (See Section 3) will be ≈21 MHz. Since the linac frequency must be a harmonic of the cyclotron frequency, it seems reasonable to assume that the initial linac stages will have a frequency of ≈84 MHz. As a consequence, the beam will need to be subharmonically bunched to match the cyclotron rf frequency. The subharmonic bunching was discussed during the first meeting of the ISOL Technical Task Force[2], and it was agreed that this would not be a serious technical issue.

For the purposes of this analysis, it is assumed that the pre-accelerator system will be basically that proposed by ANL[2] as delineated in Table 1. (Note that the empirical formula of Baron et al[3] has been used to determine the stripping energy, ≈17 MeV/u, appropriate to achieve Q/A≈1/3 for uranium ions.)

| Major Element | Voltage (MV) | Beam Energy (MeV/u) | Q/A |
|---|---|---|---|
| ECR Ion Source | | | 1/12 |
| 84 MHz CW RFQ | 1.8 | | 1/12 |
| 84 MHz NC IH Linac | 5.3 | | 1/12 |
| SC Low-energy Linac | 210 | ≈17 | 1/12 |
| SC Mid-energy Linac | 40 | ≈30 | 1/3 |

**Table 1.** Proposed 30 MeV/u (Q/A≈1/12) pre-accelerator system.

Based on the linac system given in Table 1, an analysis of the energies achievable for the lighter ions was made. Figure 1 shows a possible acceleration plan for a linac system to produce 30 MeV/u U for ions with Q/A=1/3. Figure 2 and Figure 3 show the possible energies achievable for $^4$He and $^1$H, given the linac layout of Figure 1. $^4$He energies of ≈90 MeV/u (360 MeV total) and proton energies of ≈145 MeV would be achievable with this system.

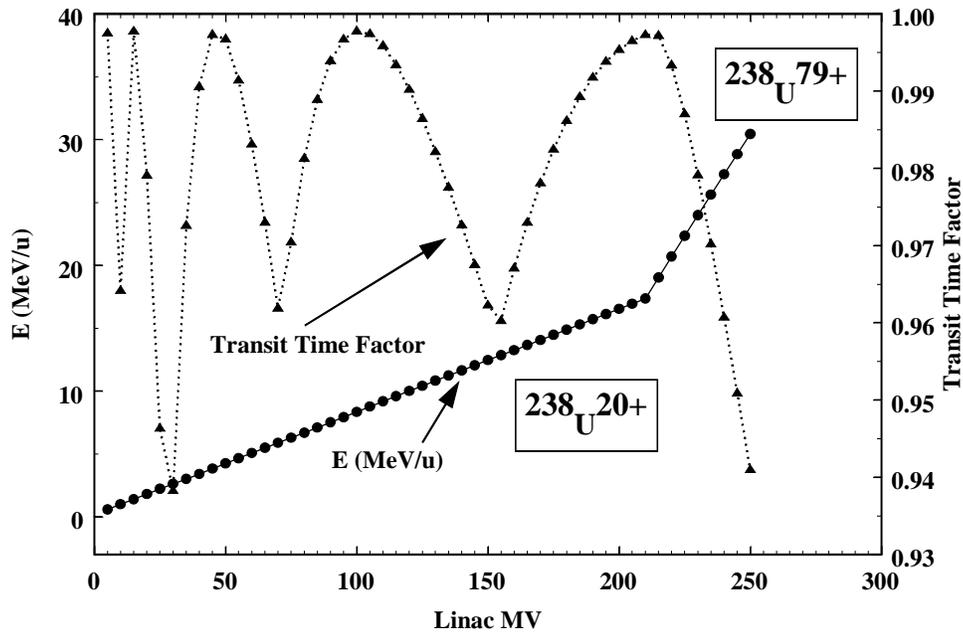

**Figure 1.** Energy and transit time factor for 30 MeV/u U.

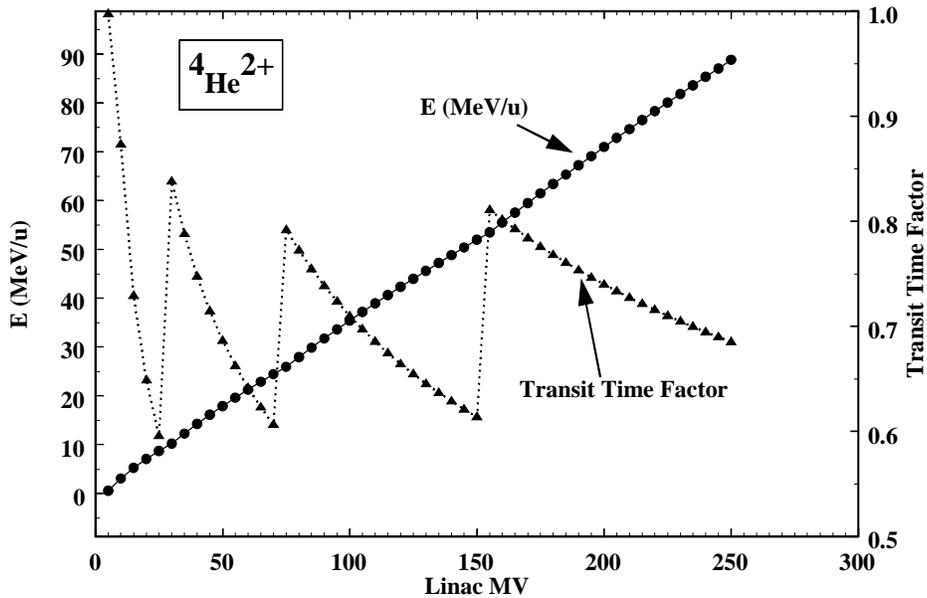

**Figure 2.** Energy and transit time factor for 90 MeV/u $^4$He.

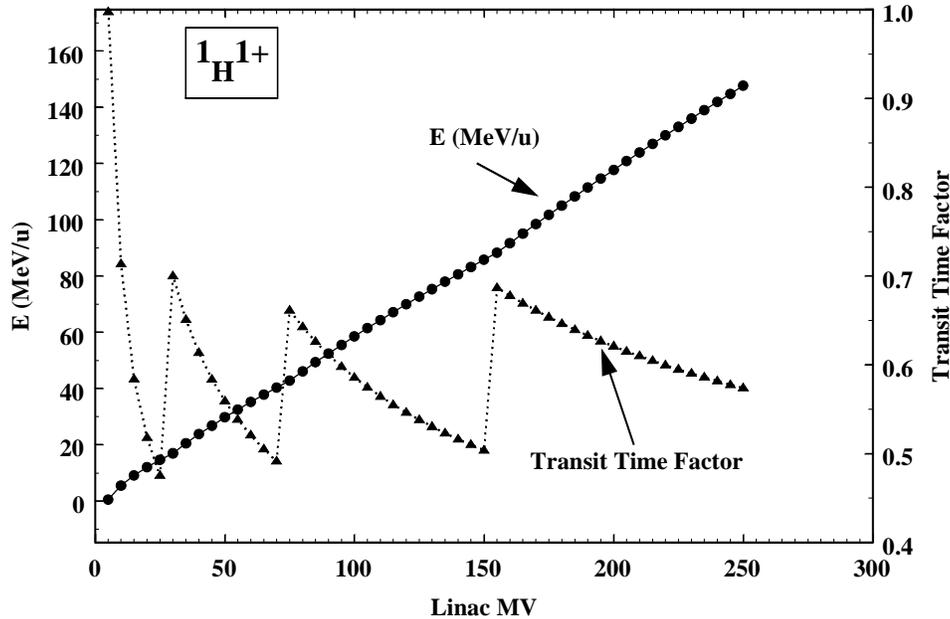

**Figure 3.** Energy and transit time factor for 140 MeV/u $^{1}$H.

## 2.2 Pre-Accelerator Stripping System

For injection into the Separated Sector Cyclotron, the pre-accelerator system must produce ions of 30 MeV/u and a Q/A≈1/3 with current levels appropriate for 100 kW of beam power at 400 MeV/u.

Beam current levels of Q/A≈1/3 ions sufficient to meet the 100 kW beam power criterion may be obtained directly from present-day ECR ion sources for ions lighter than argon. Ar and heavier ions will be accelerated to the energy required to strip the beam to Q/A≈1/3. Hence, the lightest ion that must be stripped is Ar. (Note that throughout this analysis the empirical formula of Baron et al[3] has been used to determine the stripping energy appropriate to achieve Q/A≈1/3.) The beam energy at stripping must vary from Ar at 1.7 MeV/u to Bi at 13 MeV/u. Above Bi, only Th and U require stripping at about 17 MeV/u. Each beam will be stripped only once, however, due to the different energy necessary for each element, three strippers will be required at different positions along the linac.

Figure 4 shows the possible locations along the linac of the three strippers required to reach Q/A≈1/3 for all ions. The first stripper, positioned after 75 MV of linac acceleration, is a point at which Ar through Xe would be accelerated to 1.7 through 8 MeV/u respectively. In the proposed scheme, the Ar beam will have the largest velocity excursion relative to the optimal design value for U. The Ar velocity will be 48% below that for uranium at the 75 MV point in the linac. This will decrease the transit time factor, but the excess accelerating gradient available will easily mitigate this issue. The second stripper, positioned at the ≈140 MV point of the linac, would be used for Xe through Bi. The final stripper, positioned at the ≈204 MV point in the linac, will be used for Th and U.

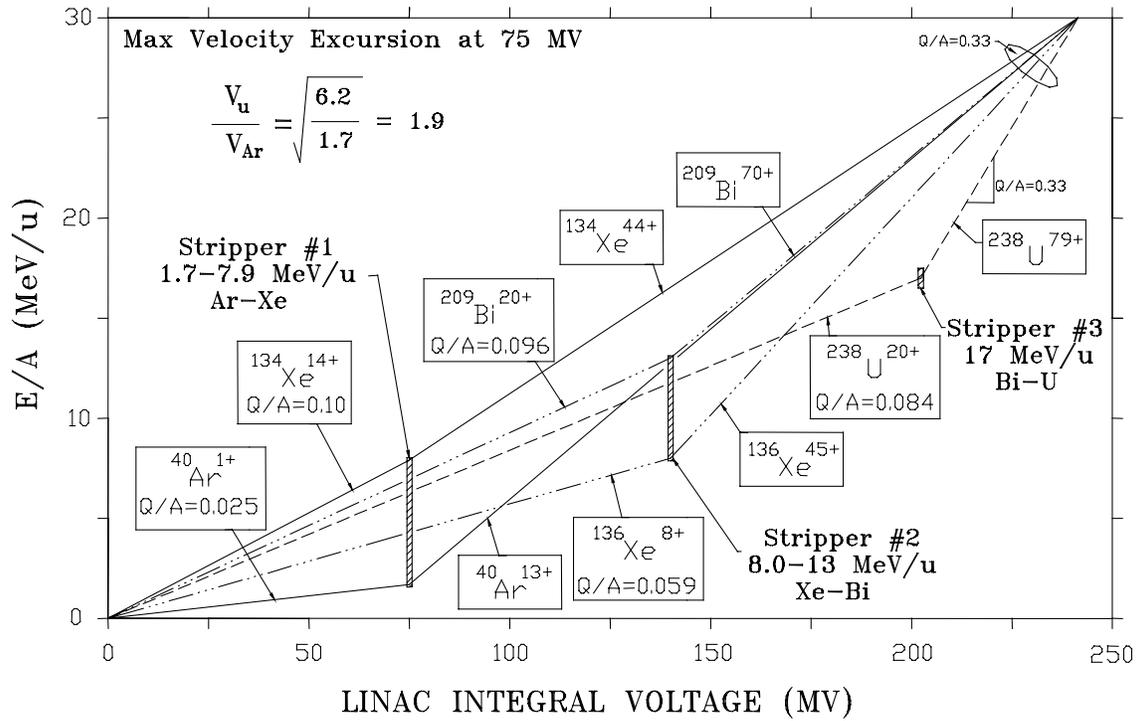

**Figure 4.** Stripper foil locations for representative beams.

### 2.3 Pre-Accelerator Longitudinal Injection Requirements

The longitudinal acceptance of the Separated Sector Cyclotron ($\approx 40\ \pi$ keV/u ns) is much larger than the longitudinal emittance ($\approx 2\ \pi$ keV/u ns) from the linac and smaller than the longitudinal acceptance of the linac ($\approx 200\ \pi$ keV/u ns). However, some manipulation of the longitudinal phase space will be required to match the longitudinal emittance of the linac to that required for the Separated Sector Cyclotron.

The cyclotron acceptance at 30 MeV/u and 21 MHz requires a beam with $\Delta t = \pm 1.3$ ns $= \pm 10^{o}$ and $\Delta E/E < \pm 0.1$ % providing a maximum longitudinal acceptance of $\approx 40\ \pi$ keV/u ns. The bunch length is set to be as long as possible compatible with the $3^{rd}$ harmonic rf flat-top system to reduce the energy spread from the longitudinal space charge force. The linac longitudinal emittance at 30 MeV/u assuming a 168 MHz linac structure is estimated to be $2\ \pi$ keV/u ns with $\Delta t = \pm 1.6^{o} = \pm 0.026$ ns and $\Delta E/E = \pm 0.26$ %.[4] By modification of the longitudinal dynamics in the last sections, the linac energy spread could be doubled by, for example, shifting the synchronous phase to the unstable region or by bunch rotation in a mismatched bucket. Under these circumstances, the linac would provide a longitudinal emittance of $2\ \pi$ keV/u ns but with $\Delta t = \pm 0.8^{o} = \pm 0.013$ ns and $\Delta E/E = \pm 0.52$ %. The beam would then drift through $\approx 35$ m and a 0.77 MV, 84 MHz rf system would reduce the energy spread as shown in Figure 5. The beam would then travel an additional $\approx 20$ m before being injected into the Separated Sector Cyclotron at which point the beam bunch length will have increased by $\approx 10$ % to the desired $\pm 1.3$ ns ($\pm 10^{o}$).

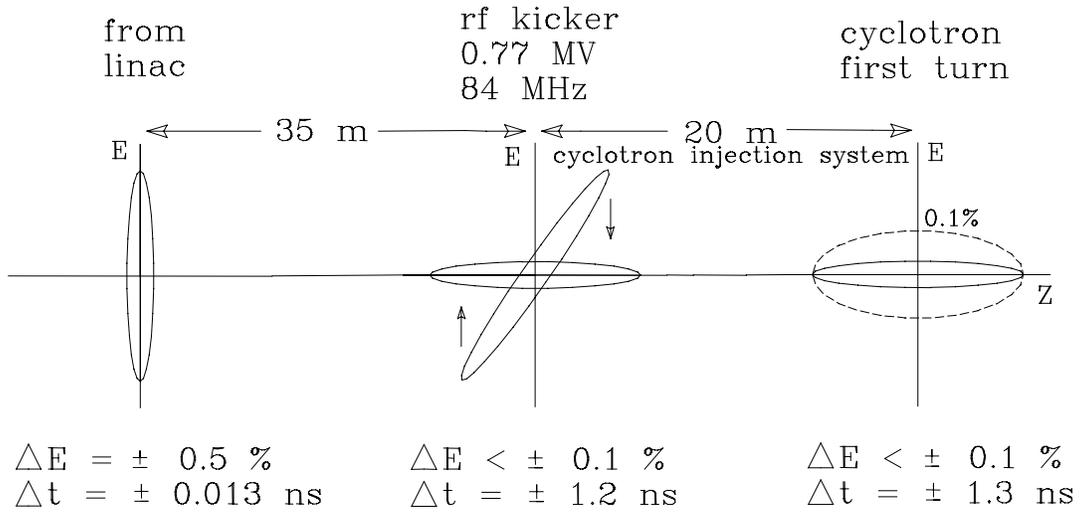

**Figure 5.** Bunch rotation scheme for injection into the cyclotron. The dashed ellipse represents the cyclotron longitudinal admittance.

Detailed linac simulations will be required to verify that the proposed modification of the linac longitudinal dynamics will achieve the desired results. An alternative bunch rotation scheme is shown in Figure 6. The additional rf system and drift length used in this scheme are not expected to be needed, but are presented because they provide a clear solution that does not appreciably impact the facility cost. In this case, beam with an energy spread of 0.26 % energy would drift ≈13 m at which point a 1 MV, 672 MHz rf system would be used to increase the energy spread to 0.5 %. The beam would then follow the same drift and rf kicker procedure as described above.

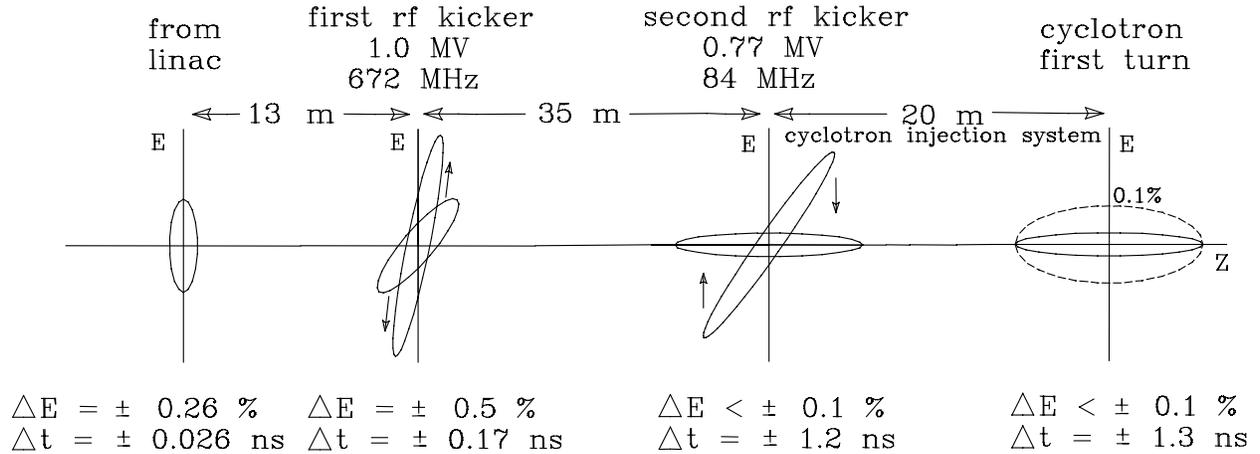

**Figure 6.** Alternative bunch rotation scheme for injection into the cyclotron. The dashed ellipse represents the cyclotron longitudinal admittance.

# 3   Separated Sector Cyclotron

An analysis of a Separated Sector Cyclotron has been completed with details provided in the sections following. The design largely follows that of the PSI Ring Cyclotron which has an extraction efficiency of 99.99%. The studies provided below predict the same high efficiency extraction for the proposed Separated Sector Cyclotron, and with a Q/A acceptance of $1/3\pm10.1\%$ will present no limitation on fragment production rates.

A 3D model is given in Figure 7 where the six laminated sector magnets, the main magnet coils, the four primary rf resonators, and the single $3^{rd}$ harmonic resonator are shown. The mechanical detail of the steel assembly for one of six sector magnets for the Separated Sector Cyclotron is provided in Figure 8.

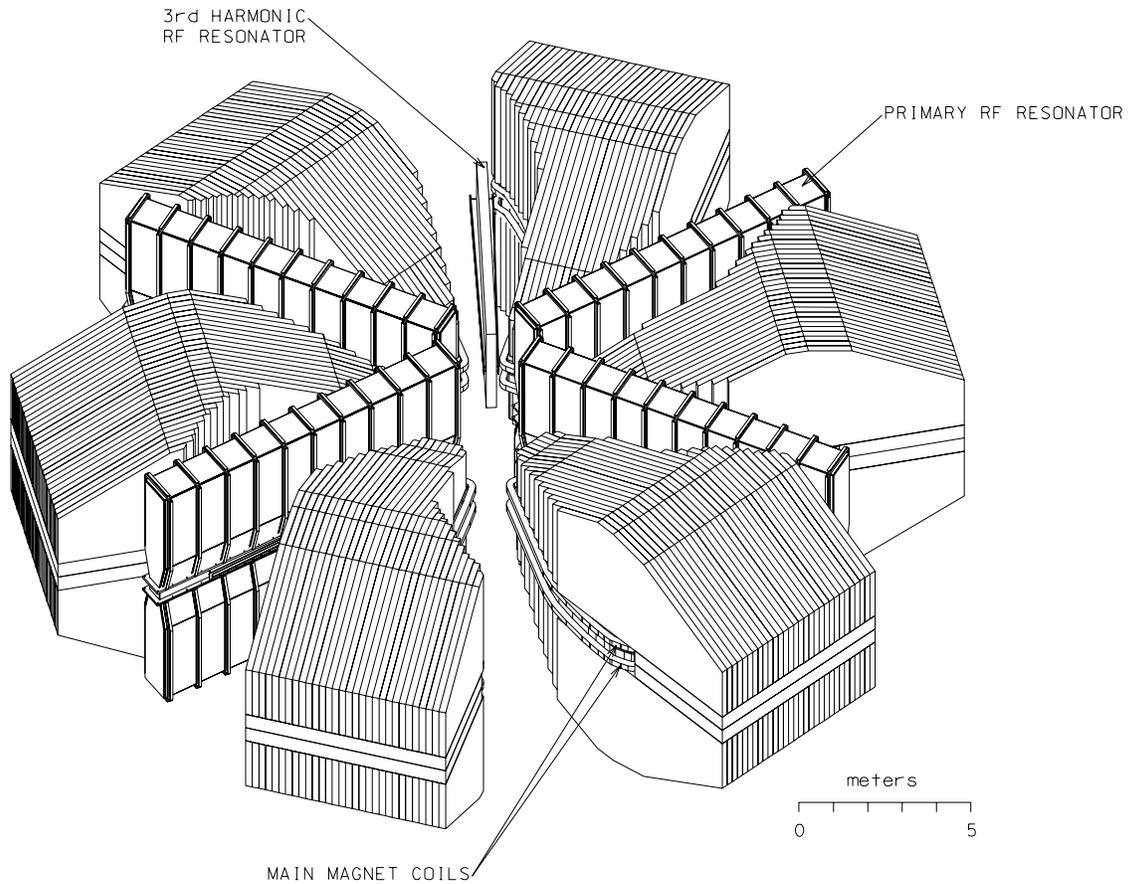

**Figure 7.** Three-dimensional model of the Separated Sector Cyclotron showing the laminated sector magnets, the main magnet coils, the four primary rf resonators, and the single $3^{rd}$ harmonic resonator.

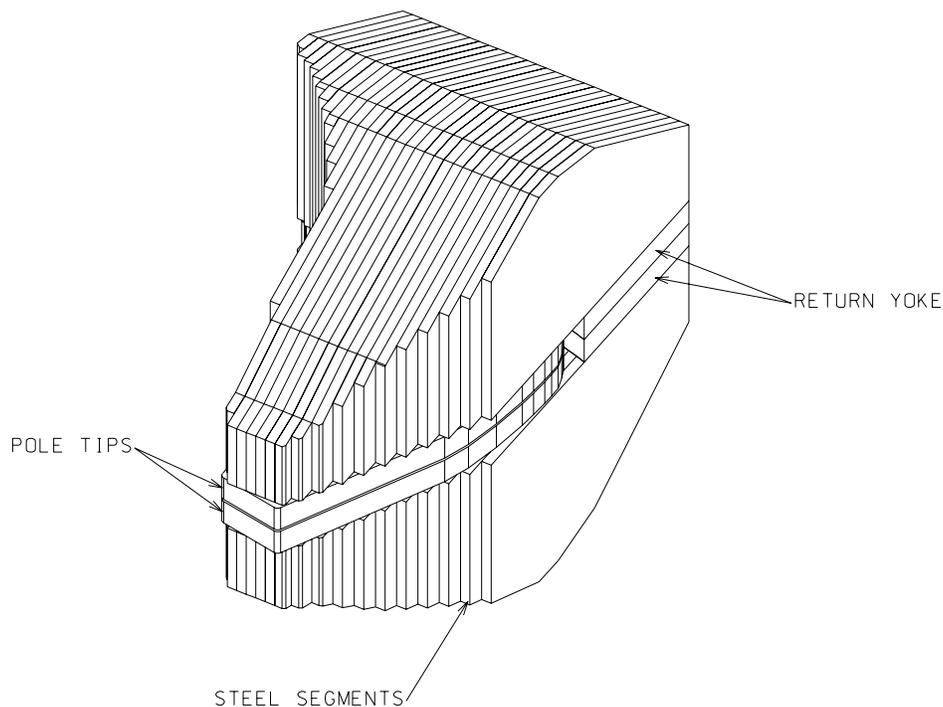

**Figure 8.** Mechanical detail of the steel configuration for one of six sector magnets for the Separated Sector Cyclotron.

### 3.1  Injection

The injection system will provide beam matching and transport from the longitudinal phase space matching elements after the pre-accelerator to the injection point of the Separated Sector Cyclotron. Since the injection path spans regions appropriate for simple beam transport models as well as regions within which the magnetic field of the Separated Sector Cyclotron must be appropriately included, two computer codes were used; DIMAD[5] for the simple beam transport region and CYCLONE[6] for the region near and within the Separated Sector Cyclotron. The general injected beam parameters required for the Separated Sector Cyclotron are given in Table 2. The general layout of the injection system is given in Figure 9 with the interior region delineated in Figure 10.

| Parameter | Value |
|---|---|
| $\varepsilon_{x,y}$ | 1.64 π mm-mrad |
| $\varepsilon_n$ | 0.4 π mm-mrad |
| Energy | 30 MeV/u |
| Q/A | ≈1/3 |
| $\Delta p/p$ | ±0.05 % |
| Cavity Voltage at Injection | 400 kV |

**Table 2.** General Separated Sector Cyclotron injected beam parameters.

The proposed injection system consists of five quadrupoles, six dipoles, and one electrostatic deflector. Quadrupoles Q1 through Q4 and dipole B1 provide the initial matching of the beam from the linac to that required for the cyclotron. Dipole magnets B2 and B3 together with quadrupole Q5 provide the required bending and focusing for the beam inside the cyclotron inner radius. Dipole B4, a sector magnet pole tip extension, will be used to shield the beam from the strong vertically defocusing field of the sector magnet's coil, and dipole B5 is a magnetic channel inside the cyclotron sector magnet. Finally, the septum magnet B6 and electrostatic deflector E1 will place the injected beam onto the desired acceleration orbit. The specifications for these elements are given in Table 4.

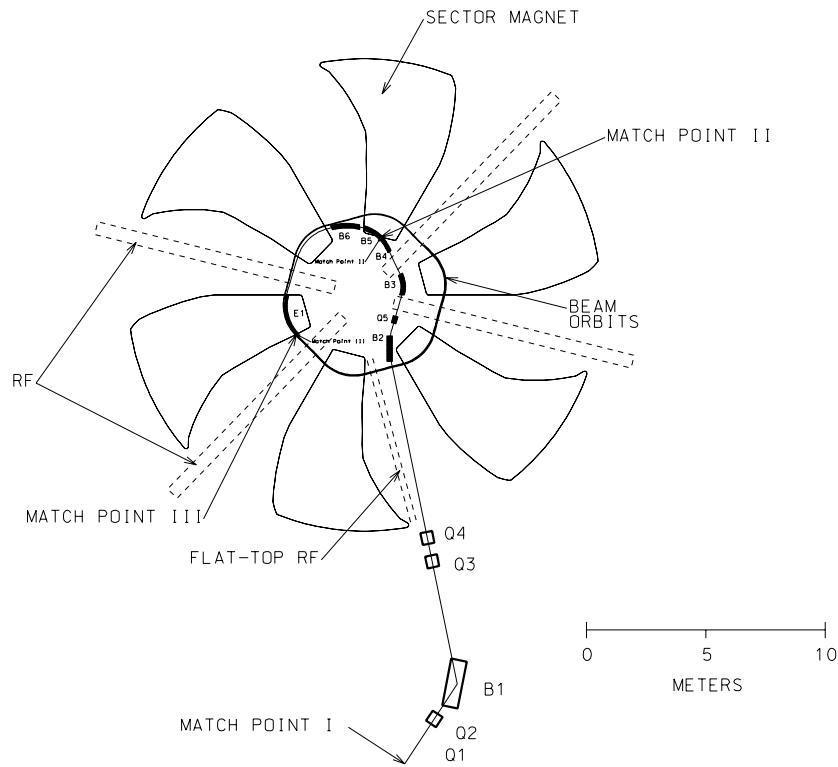

**Figure 9.** Layout of injection system for Separated Sector Cyclotron.

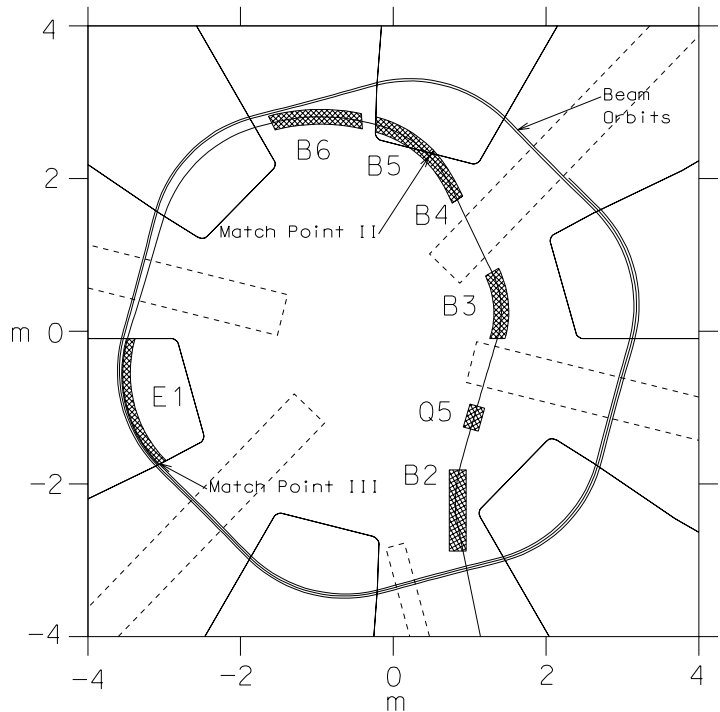

**Figure 10.** Interior region of injection system for Separated Sector Cyclotron.

The injection system was conceptually divided into three regions by specifying three matching points at which the beam machine functions were specified. (See Figure 9 and Figure 10.) Match Point I is at the entrance to the injection system, Match Point II is at an intermediate position upstream of which DIMAD provides an appropriate model and downstream of which CYCLONE is required to adequately model the Separated Sector Cyclotron environment, and finally, Match Point III is the point where the injected beam joins the design acceleration orbit of the cyclotron.

The $\beta_{x,y}$ and $\alpha_{x,y}$ machine functions at Match Point III were derived from the eigen ellipses for the cyclotron design orbit. The $\beta_{x,y}$ and $\alpha_{x,y}$ machine functions at Match Point II were found by backward tracking of the eigen ellipses of Match Point III to Match Point II using the program CYCLONE. The resulting horizontal and vertical beam envelopes are shown in Figure 11. The dispersion function, $\eta_x$, required at Match Point II was determined by the backward tracking of particles with different momenta ($\Delta p/p = \pm 0.05$ %) and appropriate radial positions within the cyclotron to Match Point II. This procedure results in values for $\eta_x$ as given in Figure 12. Finally, the machine functions for Match Point I were derived by backward tracking with the program DIMAD from Match Point II to Match Point I with the magnetic elements adjusted to produce an achromatic beam with an upright ellipse at Match Point I.

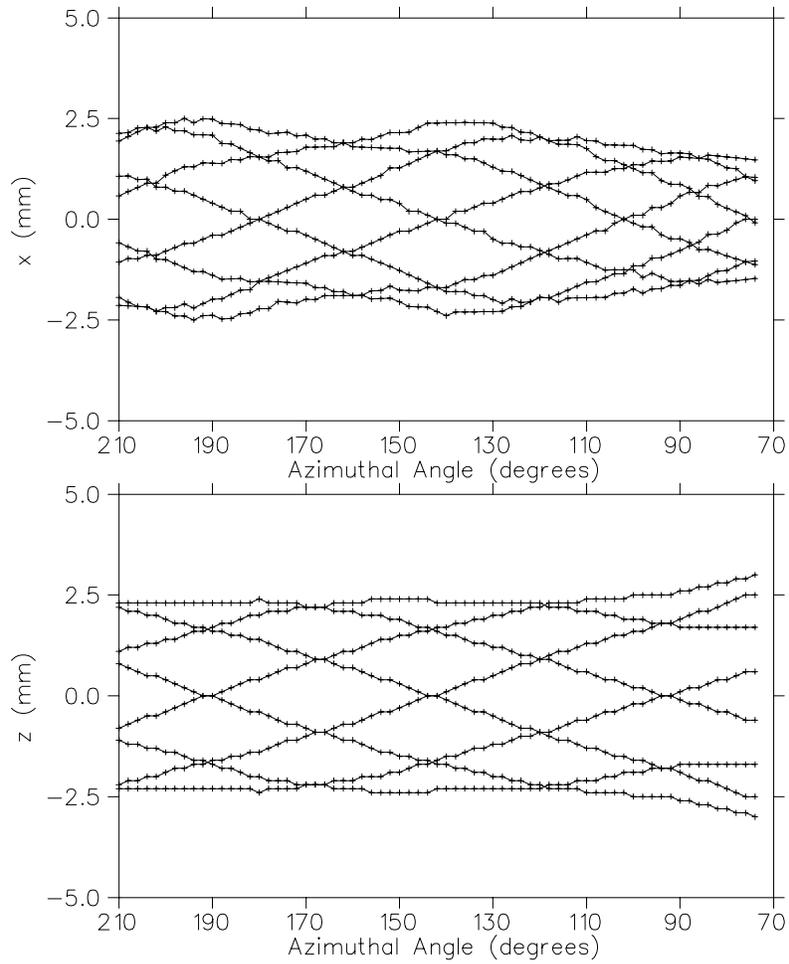

**Figure 11.** Horizontal and vertical plane results of backward tracking of particles from Match Point III to Match Point II with a beam emittance of 1.64 π mm-mrad.

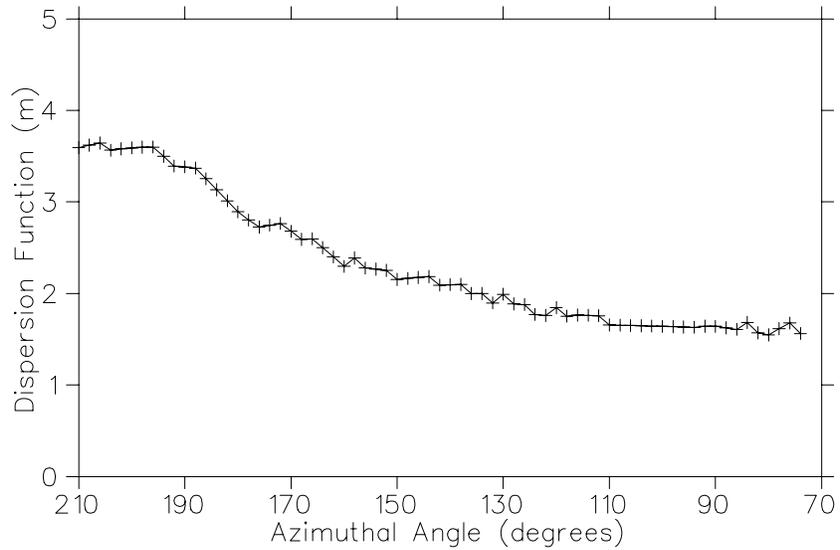

**Figure 12.** $\eta_x$ as a function of position traveling from Match Point III to Match Point II.

These analyses provide the beam function requirements at the three match points as given in Table 3. DIMAD was used to determine the hardware values required to obtain the machine functions at Match Point I from those of Match Point II with the beam envelopes over this region given in Figure 13.

| Match Point | $\beta_x / \beta_y$ (m) | $\alpha_x / \alpha_y$ | $\eta_x$ (m) / $\eta'_x$ |
|---|---|---|---|
| I | 6.56 / 5.44 | 0 / 0 | 0 / 0 |
| II | 1.32 / 5.56 | 0.18 / 1.32 | 1.56 / -0.55 |
| III | 2.97 / 3.70 | 0.403 / -0.34 | 3.29 / 0.0 |

**Table 3.** Machine function requirements at the injection system match points.

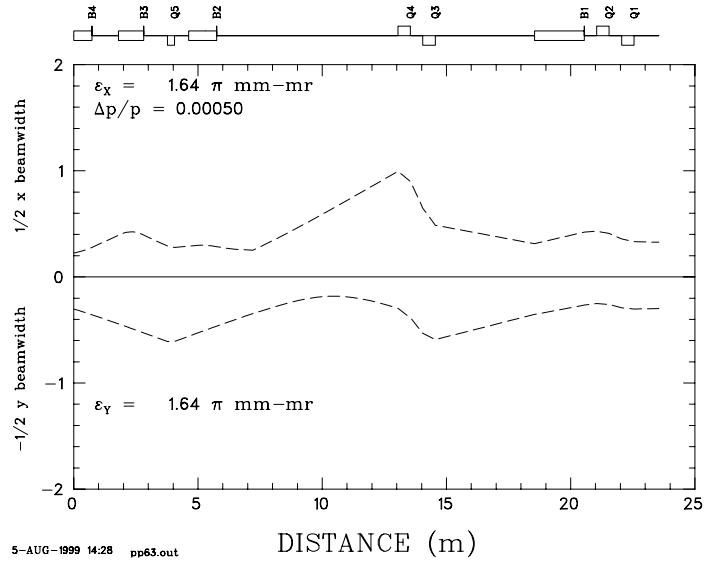

**Figure 13.** Beam envelopes from DIMAD over region from Match Point II to Match Point I.

| Element Label | Type | Strength | Effective Length (m) |
|---|---|---|---|
| Q1 | Quadrupole | 1.3 T/m | 0.5 |
| Q2 | Quadrupole | 1.4 T/m | 0.5 |
| Q3 | Quadrupole | 3.0 T/m | 0.5 |
| Q4 | Quadrupole | 3.1 T/m | 0.5 |
| Q5 | Quadrupole | 2.4 T/m | 0.28 |
| B1 | Dipole | 9.4 kG | 2.0 |
| B2 | Dipole | 12.5 kG | 1.2 |
| B3 | Dipole | 20.0 kG | 1.0 |
| B4 | Sector pole tip magnet | 7.5 kG | 0.75 |
| B5 | Magnetic Channel | 3.0 kG | 0.75 |
| B6 | Magnetic Septum | 10.0 kG | 1.15 |
| E1 | Electrostatic Septum | 30.0 kV/cm | 1.75 |

**Table 4.** Magnetic and electrostatic elements required for the proposed Separated Sector Cyclotron injection system.

The hardware required for the injection system is listed in Table 4. None of these elements are expected to pose significant technical challenges. The center-to-center beam separation between the injected beam and that of the next cyclotron orbit is 12.6 cm at the B6 exit point where the beam full width is ≈1 cm in the horizontal plane and ≈0.5 cm in the vertical plane. The design for B6 provides for a septum thickness of ≈3.5 cm providing ≈4 cm beam clearance at the exit point. Similarly, the center-to-center beam separation between the injected beam and that of the next cyclotron orbit is ≈2.7 cm at the E1 entry point where the beam size is similar. The design for the E1 specifies a septum thickness of ≈0.05 cm providing ≈0.8 cm beam clearance at that point. Hence, the injection system should be effectively 100% efficient.

**3.2 Main Cyclotron**

3.2.1 Q/A Range

The basic premise of the design for the Separated Sector Cyclotron was to keep it as simple as possible. In this regard, an initial analysis[1] proposed that the injected ions have a Q/A = 1/3 ±1%. Since that time, a more detailed evaluation has been undertaken to determine what Q/A range would reasonably be required and what concomitant trim coil parameters would be necessary. Figure 14 presents the deviation from Q/A = 1/3 as a function of atomic number (Z) for all stable isotopes.

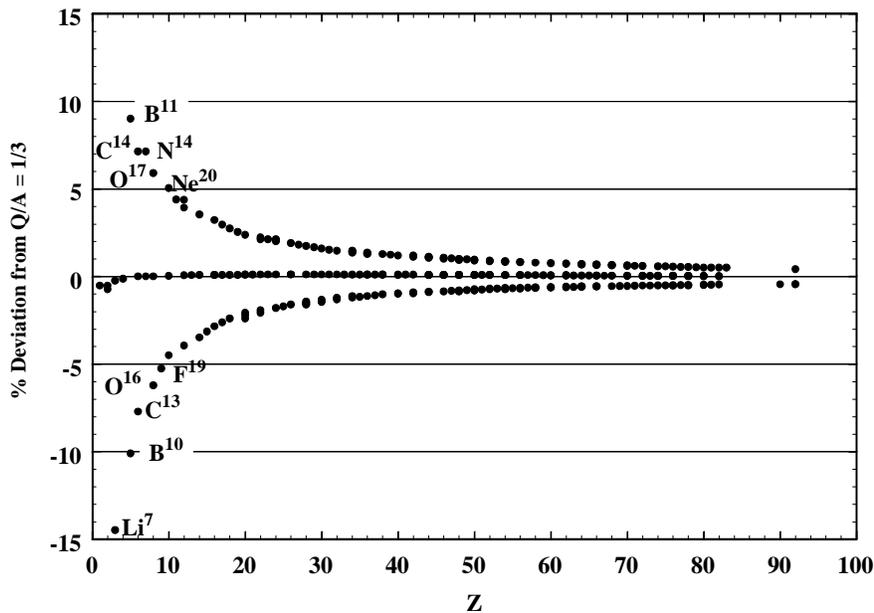

**Figure 14.** Deviation from Q/A = 1/3 as a function of atomic number for all stable isotopes.

With appropriate selection of Q, a Q/A range of ±5.1% would include all possible stable isotopes excepting $He^4$, $Li^7$, $B^{10}$, $C^{13}$, $O^{16}$, and $F^{19}$ on the negative side and $B^{11}$, $C^{14}$, $N^{14}$, and $O^{17}$ on the positive side. The consequences on the radioactive beam production rates assuming a Q/A range of ± 5.1% are given in Figure 15. Even in this circumstance, the production rate is only affected by a factor of 5 or less for a few isotopes.

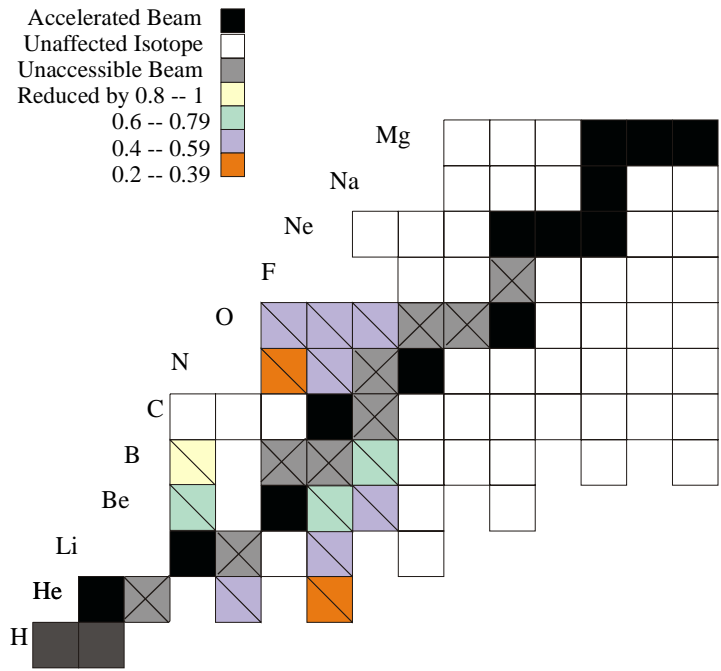

**Figure 15.** Production rate effect of choosing a Q/A bandwidth of ±5.1% affects production of 13 isotopes by a factor of 5 or less.

However, a Q/A range of ± 10.1 % would include all possible stable isotopes excluding only $^4$He and Li$^7$. Only $^4$He and $^7$Li will not be included within this range; proton, deuteron, and helium beams would be provided by utilizing H$_3^{1+}$ and DH$^{1+}$ molecules and $^3$He$^{+1}$ ions respectively.

**<u>A Q/A range of 1/3±10.1% will cause no loss of fragment intensity. As a consequence of this analysis and the corresponding achievable trim coil range, the Separated Sector Cyclotron Q/A range will be 1/3 ±10.1%.</u>**

3.2.2   Trim coils

To determine the strength of trim coils necessary to fit the range of charge to mass ratios 1/3 ± 10.1 %, the following procedure was used.

1. **<u>The Base Isochronous Field was determined.</u>** The isochronous field required for the Q/A=1/3 and 400 MeV/nucleon (Base Isochronous Field) case was determined by a TOSCA model of a sector magnet that provided the approximately correct field. This field corresponds to a total of ≈100,000 ampere-turns per coil. The TOSCA field map was then modified to obtain the radial profile required for isochronicity. In the actual cyclotron, this field shape would be obtained to a close tolerance by appropriate shimming of the magnet.
2. **<u>The Variation of the Base Isochronous Field was determined.</u>** To obtain the change in the magnetic field as a function of the main coil current, the TOSCA model was run at two different excitations and the fields subtracted to obtain the derivative of the magnetic field with respect to the main coil excitation. It was then assumed that the base magnetic field for different Q/A values can be obtained from the sum of the isochronous field for Q/A=1/3 and the derivative field multiplied by the coil excitation difference.
3. The isochronous fields for other Q/A ratios correspond to simply multiplying the Base Isochronous Field by the corresponding scaling factor for the Q/A variation.

4. A least square fit of the difference field obtained in Item 2 (by changing the main coil current to fit the field required from Item 3) was then done. The residual error after fitting the Q/A=0.366 field is shown in Figure 16.

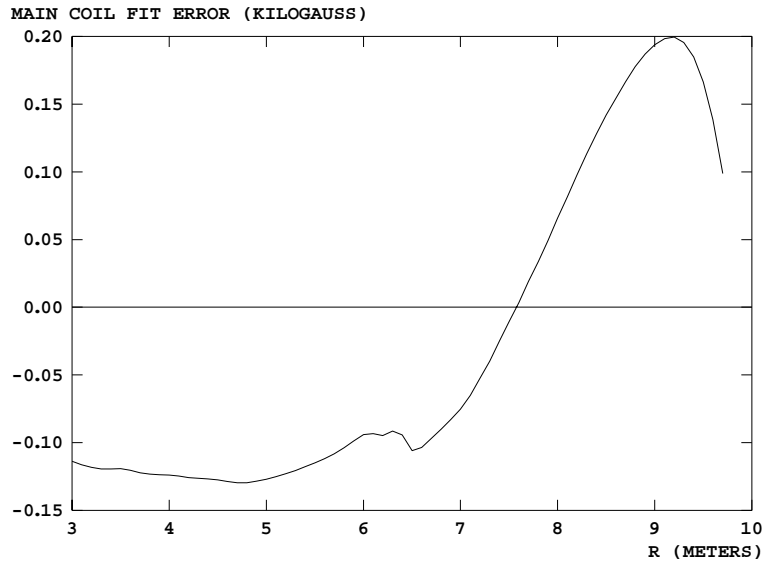

**Figure 16.** Average field difference between the TOSCA prediction and the isochronous field for Q/A = 0.366 as a function of radius.

From Figure 16, it can be seen that the TOSCA fields show an excess field of approximately 0.2 kG in the region of large radii, and a defect of 0.1 kG at small radii. This can be interpreted as the effect of iron saturation causing the magnetic circuit to become more "air core" like and hence, the path reluctance to be less dependent on the air gap. Under this circumstance, the coil field would show a greater increase at larger radii where the path reluctance is smaller; a result compatible with Figure 16.

Because of the small magnet gap, it was decided not to locate the trim coils in the gap space, but rather to place them inside the steel of the pole tip. To avoid increasing the reluctance, the coils were oriented vertically in 1 cm (width) by 10 cm (height) slots centered at a position 25 cm from the median plane as shown in Figure 17.

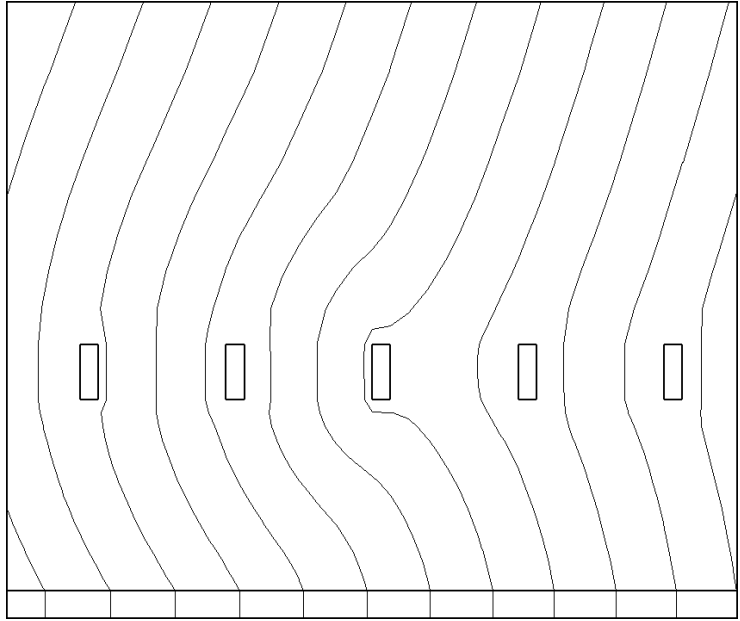

**Figure 17.** A vertical cross section of the POISSON model showing holes in the pole tip for the trim coils with the median plane given by the bottom horizontal line.

Though not fully saturated, the field in the steel is nonlinear, and ultimately will require a non-linear successive approximation technique to find the total field produced by the trim coils. However, in order to obtain a first estimate of the trim coil power requirements, linearity has been assumed. A 2D POISSON model was created consisting of a simplified azimuthally symmetric magnet with circular trim coils. The return path for the trim coils will either be on the outer radius of the magnet (for those at larger radii) or the inner radius (for those at smaller radii). This model gives an approximation to the 3D model that will later be constructed with TOSCA or a similar program. To simplify the POISSON grid, the holes were represented by regions of 2 cm x 6 cm instead of 1 by 10. A small section of the POISSON model is shown in Figure 17.

Figure 18 shows the field produced by a trim coil in the middle of the pole tip at a trim coil excitation of 6000 ampere-turns. Under this condition, the trim coil field produced is approximately 500 gauss.

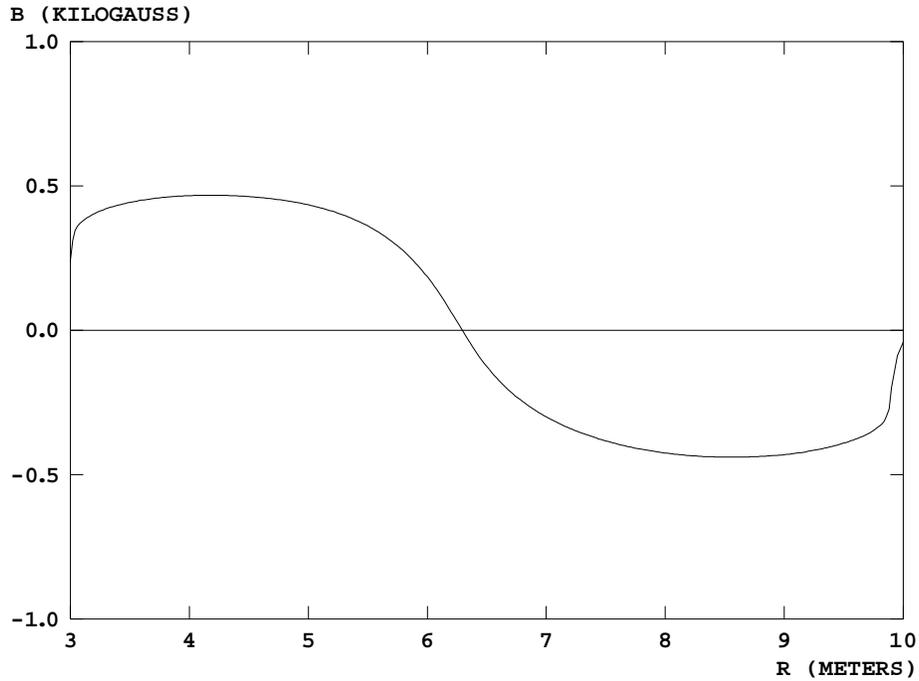

**Figure 18.** Mid-plane average field produced by one trim coil with an excitation of 6000 ampere-turns.

Using a least square procedure, the currents in all 14 trim coils used in the calculation and the main coil excitation were determined. The fit required currents of less than 1000 ampere-turns for most of the coils (less than 5000 ampere-turns for the last three at large radii and for the two innermost). The average field error in the resulting fit can be seen in Figure 19. The amplitude variation in the middle radii is probably attributable to irregularities and noise in the TOSCA grid. Even so, for most radii, the error is less than 2 gauss.

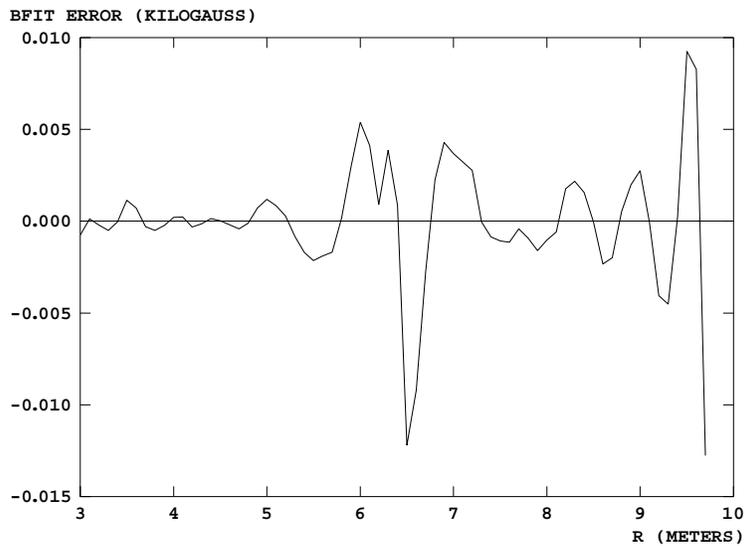

**Figure 19.** Resulting average field error in the fit to the isochronous field for Q/A = 0.366.

Figure 20 shows the phase slip resulting from this first crude fit, and establishes that an adequately isochronized field can be obtained with a network of trim coils of the general form shown here. Note that while 14 trim coils were assumed in this analysis, the costing has been calculated assuming 30 such coils to allow for the possibility of increasing the coil number above that of this first estimate. Moreover, the set of 14 trim coils assumed in this example requires a reasonable total power of ≈300 kW.

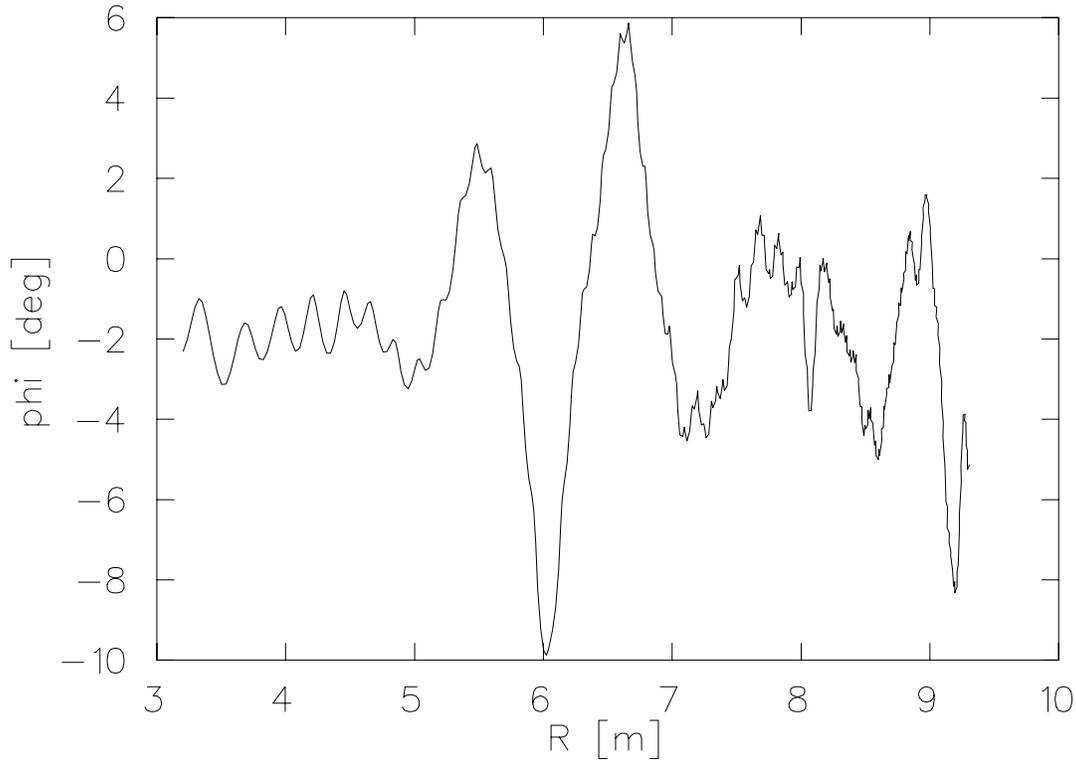

**Figure 20.** Resulting phase oscillation induced by the field error shown in Figure 19.

3.2.3   Vacuum Requirements

The attenuation of the beam at various pressures in the cyclotron was calculated using estimated cross sections for the gain or loss of an electron by an ion in the accelerated beam from collisions with residual gas molecules. The cross section estimates (applicable to Kr and heavier ions) used were those from Franzke[7] obtained by fitting the theoretical estimates of Bohr and Lindhard[8]. The charge changing cross sections were also calculated with the formulas of Betz and Schmelzer[9] that were developed for velocities up to several MeV/u. These predicted less attenuation of the beam by nearly an order of magnitude, and the more pessimistic Franzke formalism was used. The total cross section decreases with energy over the applicable range in the cyclotron, 30 to 400 MeV/u. The path length for 425 turns is 17.5 km. The residual gas was represented as $N_2$ molecules at room temperature, so the total cross section per atom was doubled to estimate cross section per molecule.

Several representative ions were compared. A transmission ratio of 0.997 was found for $^{238}U^{79+}$ ions at a 1 x $10^{-7}$ torr residual gas pressure. This corresponds to a beam loss of 170 W integrated over all energies in the accelerator for 1 particle μA injected intensity. Somewhat surprisingly, the transmission factor at the same pressure was only slightly lower for Kr and Xe ions of the same charge to mass ratio, 0.33. The predicted average charge changing cross section was roughly the same, despite some differences in the energy distribution. Both electron capture and loss made significant contributions to beam attenuation for the U beam, but the attenuation was dominated by the electron loss process for the lighter ions. A pressure of 1 x 10$^{-7}$ torr was chosen as the design goal.

The Figure 21 and Figure 22 below give the transmission and the power of the lost beam respectively, as a function of the pressure in the cyclotron for $^{238}$U$^{79+}$.

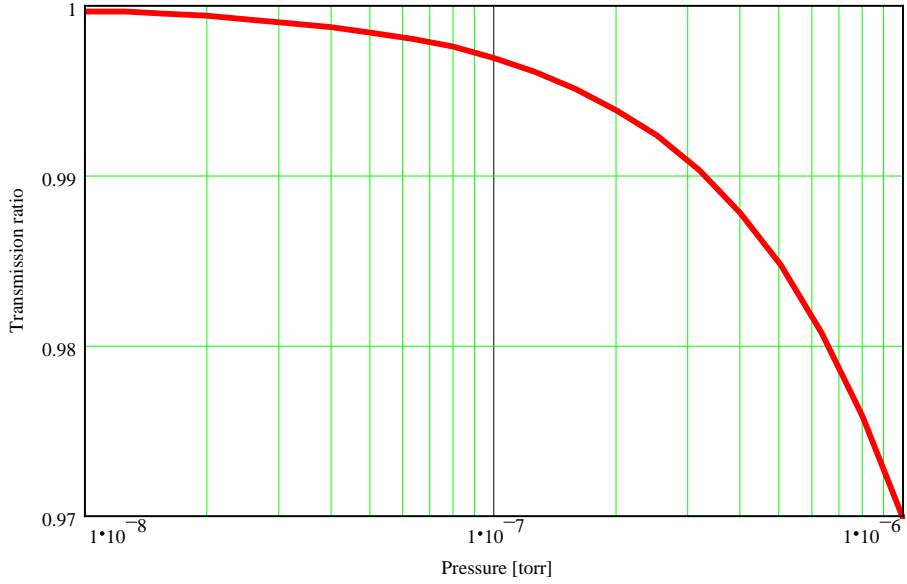

**Figure 21.** Percentage of beam transmitted as a function of vacuum pressure assuming the Franzke formalism[7].

The formulas for the capture and loss cross sections [cm$^2$/atom] given in Ref. 7 are:

$$\sigma_C = 2.0 \cdot 10^{-24} Z^{0.5} \overline{q}^{-2} \overline{q_T} (\gamma^2 - 1)^{-2} \left[ \frac{\overline{q}}{q} \right]^a$$

$$\sigma_L = 3.5 \cdot 10^{-18+X} \overline{q}^{-2} \overline{q_T} (\gamma^2 - 1)^{-0.5} \left[ \frac{\overline{q}}{q} \right]^b$$

Where $\overline{q}$ and $\overline{q_T}$ are the equilibrium charge state of the ion and the target, respectively, Z is the ion atomic number, $\gamma$ is 1+T/931.5 MeV and X=(0.71 log$_{10}$Z)$^{1.5}$. a=2 and b=-4 for high charge states $q \geq \overline{q}$, and a=4 and b=-2.3 for $q < \overline{q}$. The equilibrium charge is estimated from the formula[10]

$$\overline{q} = Z \left[ 1 - e^{-(137\beta\delta)} \right]$$

Where $\beta$ represents the ratio of the velocity to the speed of light, and $\delta$ is given by

$$\delta = 0.3443 - 0.0667 \ln(Z)$$

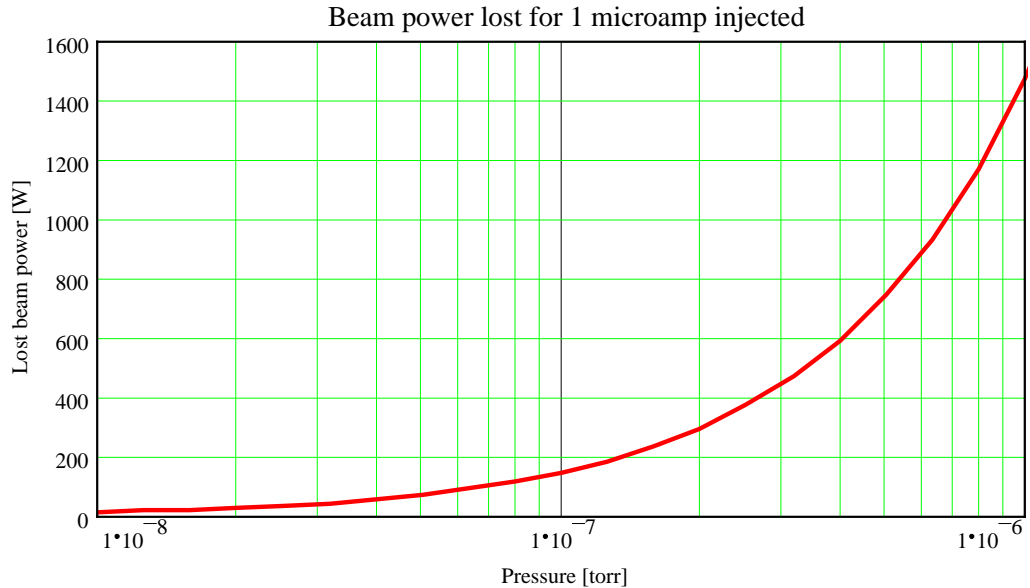

**Figure 22.** Beam power lost for an injected beam intensity of 1 particle microamp as a function of vacuum pressure following the formalism of Franzke.[7]

3.2.4 Acceleration System

*3.2.4.1 RF Overview*
The cyclotron rf resonators will be room temperature rectangular resonators fabricated from 6061-T6 aluminum. Four resonators will be used for the main acceleration system and one will be used to provide the $3^{rd}$ harmonic flat-top field. A separate amplifier/control chain will be used to drive each resonator, although the main resonators are intended to be set equally. It is envisioned that each resonator will be capable of being removed from the cyclotron by disengaging the vacuum seals and electrical connections and moving it out in the radial direction from the cyclotron. The overall design described is similar to the design of the PSI Ring Cyclotron.

*3.2.4.2 RF Design Options and Requirements*
To identify the options and performance issues associated with the rf system choices and accelerator requirements, an RF workshop was held at Michigan State University during June 3 – 7. Attending were Peter Sigg of PSI, Claude Bieth of GANIL, and T. Saito of RIKEN. The discussion focussed on the design choice of Single-Gap (SG) versus Dual-Gap (DG) structures. As the specific characteristics of these structures were discussed, many of the requirements imposed by the accelerator were also identified. These requirements included the need for flat-topping, mechanical constraints, acceptable radial voltage profiles, and peak energy gain.

In the absence of space constraints, based on a review of existing systems it was determined that either the DG or SG rf structures could supply the appropriate acceleration.

The DG structures provide greater flexibility in designing the voltage profile along the cyclotron radius, require a smaller radial footprint, require ~ ½ the voltage for the same energy gain, and include an available field free region within the accelerating electrode. These attributes come at the expense of greater complexity, and the larger azimuthal space required.

| Main RF System Requirement | Value |
| --- | --- |
| RF Frequency | 21.47 MHz |
| Cyclotron Harmonic | 6 |
| Number of Cavities | 4 |
| Cavity Phasing | 0 Degrees |
| Maximum Beam Power | 400 kW |
| Peak Voltage | 800 kV peak |
| Voltage at Injection Radius | 400 kV peak |
| Voltage at Extraction Radius | 640 kV peak |
| | |
| **Flat-Top RF System Requirement** | **Value** |
| RF Frequency | 64.41 MHz |
| Number of Cavities | 1 |
| Peak Voltage | 400 kV peak |
| Minimum Injection Voltage | 0 |
| Minimum Extraction Voltage | 0 |
| | |
| **Mechanical Constraints** | **Value** |
| Injection Radius | 3.2 m |
| Extraction Radius | 9.5 m |
| Maximum azimuth about median plane region | 0.9 m |
| Maximum azimuth above median plane region | 1.35 m |
| Maximum outer radius mechanical access | TBD |
| Minimum inner radius mechanical access | 1.4 m |

**Table 5.** Primary Separated Sector Cyclotron rf system parameters.

The SG structures are much simpler structures that require less azimuthal space. These attributes come at the expense of a greater radial space requirement to provide the required voltage profile, and for the same energy gain, twice the voltage of the DG structures. However, the required radial space may be reduced by using techniques such as folding the cavity outside of the accelerating region given the necessary available space in the azimuthal direction.

The limited azimuthal space and large structure size led to the choice of the SG structure. As a consequence, the large azimuthal spaces on the surfaces of the vacuum vessels that span the area between the rf cavities and magnets will be available for cryopumps and phase probes that would have occupied the DG field free regions. Table 5 summarizes the accelerator requirements of the RF system.

*3.2.4.3   RF Design*

The azimuthal space available for each resonator is a rectangular region 1.35 m wide, narrowing to 0.9 m at the median plane. The radial length is set by the required voltage profile, and the height is set by rf design considerations. The space within the resonator that determines the rf parameters is further reduced to allow for the one inch thick walls and six inch structural members. This reduces the azimuthal space for the rf surfaces to a rectangular region of 0.99 m wide, narrowing to 0.54 m at the median plane. The resonator shape is shown in Figure 23. For the discussion following, a Cartesian coordinate system is chosen that places the x-axis along the cyclotron radial direction, the y-axis tangent to the cyclotron azimuthal direction, and the z-axis along the cyclotron height.

Using the coordinate system described, the $TE_{1,0,1}$ mode of a pure rectangular resonator is sought. The most efficient shape for this resonator in the xz plane would be square. The losses are reduced as the y dimension is made larger to a point where the transit-time factor in the machine center offsets any gains in efficiency by requiring a larger peak resonator voltage. Since the voltage variation along the x-axis (machine radius) is unaffected by narrowing the resonator uniformly along this axis, the resonator radial dimension is readily determined analytically so as to satisfy the accelerator requirements. The y-axis dimension about the median plane is set to fit within the mechanical constraints and have a reasonable transit time factor at the injection radius. The y-axis dimension outside of the median plane area is set to be as large as allowable to reduce resonator losses. The y-axis dimension is shaped along the z-axis in a manner to clear the magnet coils. The length along the z-axis is then adjusted to achieve the required resonant frequency.

The parameters for the main resonators were found using the COSMOS/M finite element analysis system from Structural Research and Analysis Corporation with the add-on package "High Frequency Integrated Simulator for Microwave Cavities (HIFIS/MICAV) from Integrated Microwave Technologies. The results were checked against known analytical cases with similar mesh sizes and found to agree to well within 1%. In the future, it is also planned to check this analysis using MAFIA.

The flat-top cavity requires a much smaller space and operates at ½ the main cavity voltage. Since it does not pose any significant challenges with respect to drive power and can easily fit within the available space, a pure rectangular shape is proposed. The parameters were found using the analytical formulas for rectangular cavities. This design will be optimized in the future to reduce losses.

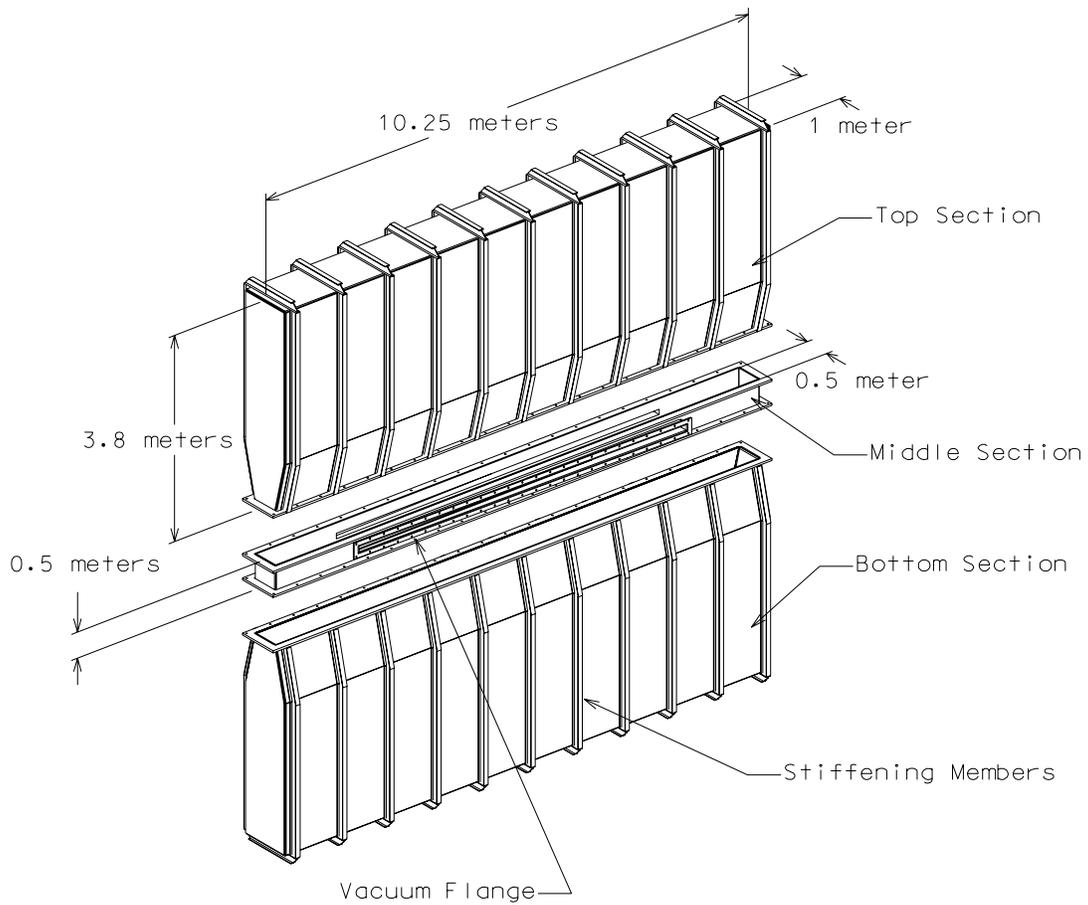

**Figure 23.** Exploded mechanical view of the resonator for the main (21.47 MHz) rf system.

Using the process and design rules described, the resonator shown in Figure 23 achieved the required rf performance parameters listed in Table 6 for the main rf resonators.

| Main RF Resonators | Value |
|---|---|
| Resonant Frequency | 21.47 MHz |
| Unloaded Q | 32,722 |
| Unloaded power at 939 kV peak | 455 kW (6061-T6 Aluminum) |
| $C_s$ | 245 pF |
| $R_s$ | 0.990 MOhms |
| $L_s$ | 224.2 nH |
| RF length along the x-axis | 10.25 m |
| RF length along the y-axis | 0.5 m to 1.0 m |
| RF length along the z-axis | 8.2 m |
| Injection point along x-axis | 1.8 m ( 3.2 m cyclotron radius) |
| Extraction point along x-axis | 8.1 m ( 9.5 m cyclotron radius) |
| Peak voltage point along x-axis | 939 kV |
| Transit-Time-Factor at injection | 0.859 |
| Transit-Time-Factor at extraction | 1.0 |
| Circuit voltage at injection | 487 kV |
| Circuit voltage at extraction | 640 kV |
| Effective voltage at injection | 418 kV (45 %) |
| Effective voltage at extraction | 640 kV (68 %) |
| | |
| **Flat-Top RF Resonators** | **Value** |
| Resonant Frequency | 64.41 MHz |
| Unloaded Q | 5,463 |
| Unloaded power at 469.5 kV peak | 780 kW (6061-T6 Aluminum) |
| $C_s$ | 152 pF |
| $R_s$ | 0.271 MOhms |
| $L_s$ | 39.6 nH |
| RF length along the x-axis | 7.30 m |
| RF length along the y-axis | 0.25 m |
| RF length along the z-axis | 2.45 m |
| Injection point along x-axis | 0.5 m ( 3.2 m cyclotron radius) |
| Extraction point along x-axis | 6.8 m ( 9.5 m cyclotron radius) |
| Peak voltage point along x-axis | 645 kV |
| Transit-Time-Factor at injection | 0.70 |
| Transit-Time-Factor at extraction | 0.96 |
| Circuit voltage at injection | 52.6 kV |
| Circuit voltage at extraction | 52.6 kV |
| Effective voltage at injection | 36.8 kV (8 %) |
| Effective voltage at extraction | 50.5 kV (11 %) |

**Table 6.** Detailed parameters for the Separated Sector Cyclotron rf system.

## 3.3 Extraction System

The primary goal of the extraction system design is the achievement of high-efficiency, single-turn extraction. I.E., the last turn must be separated from the previous turns by an amount sufficient to provide space for an electrostatic deflector septum with a typical thickness of ≈0.25 mm and error-driven beam position variations.

The effects that will reduce the turn separation are:
1. **Beam RF Phase Width** A beam with a broader rf phase width will also have a concomitant larger energy spread that will translate to an increase in the beam's radial width and hence, reduce the turn separation. In a single frequency rf system, the beam width and energy spread correlation is caused by the sinusoidal shape of the rf voltage. With the introduction of an additional rf harmonic ($3^{rd}$), the beam width and energy correlation may be dramatically reduced over a significant range of rf phase about the peak voltage point.
2. **Injected Beam Energy Spread** The energy spread of the injected beam even in the circumstance of a small energy spread and even if dispersion matched at injection, will still induce a small turn width increase. Hence, this effect will reduce the turn separation at the extraction position.
3. **RF Voltage Stability** A variation in rf voltage will result in a variation of the average turn position. Again, this will also reduce the turn separation at the extraction position.
4. **Magnet Current Variation** Variation in the magnet power supply current will result in a variation of the average magnetic field. This effect will induce rf phase excursions of the ions around the design phase. If these phase excursions become too large, they will exceed the phase width over which the phase energy correlation has been reduced (Item 1), and as a consequence the beam energy spread will be increased. Again, this will result in a beam with an increased radial dimension and hence, reduced turn separation.
5. **Space Charge** The charge in the bunch will produce an electric field that will increase the energy of the ions at the head of the bunch and reduce the energy of the ions at the tail. This increased energy spread will result in an increased radial beam width that will also reduce the turn separation.
6. **Radial-Vertical Coupling** Under the condition that the motion in the radial and vertical planes are coupled, the vertical oscillations will cause a similar motion in the radial or extraction plane, and hence reduce the turn separation.

On the other hand, the following elements may be used to increase the turn separation.
1. The radius gain per turn due to acceleration may be expressed as:
$$\frac{\Delta R}{R} = \frac{\Delta E}{E} \cdot \frac{\gamma}{\gamma+1} \cdot \frac{1}{\nu r^2}$$
   Where: R = radial position, E = energy, νr = radial tune.
   From the above equation, increasing R would increase the turn separation.
2. The above equation also demonstrates that a large gain in the turn separation would be possible by reducing νr.
3. An oscillation may be induced by, for example, injecting off center that will cause the beam to precess, and thereby increase the turn separation given a centering error of the appropriate amplitude and phase.

4. Reduction in the injected emittance would also increase the clear region between turns.

The increase in turn separation provided by increasing the machine radius has already been pursued as far as reasonable. Although, a decrease in the injected transverse emittance would have a positive effect on extraction efficiency, in the spirit of exploring a robust design, the analysis explored the effect of **increasing** the injected emittance. The magnet current variation analysis will be the subject of future work. It will largely follow that given for the rf voltage variation (Section 3.3.3), is estimated to have a tolerance of $\approx 10^{-5}$, and is not anticipated to be a significant technical issue.

### 3.3.1  νr Effect

Reducing νr provides the possibility of increasing the turn separation by extracting at a lower value of νr. This can be achieved by choosing the radial position where the magnetic field has fallen below the isochronous field. The choice of this position must, however, be balanced against the phase slip associated with a non-isochronous field. The closed orbits for two energy values that correspond to an increment equal to the energy gain per turn were calculated. From these values, the radius increase at the azimuthal angle of the electrostatic deflector entrance was determined. This procedure was repeated for several energies to determine the expected radius gain as a function of the final energy. Shown in Figure 24 are the results of this computation.

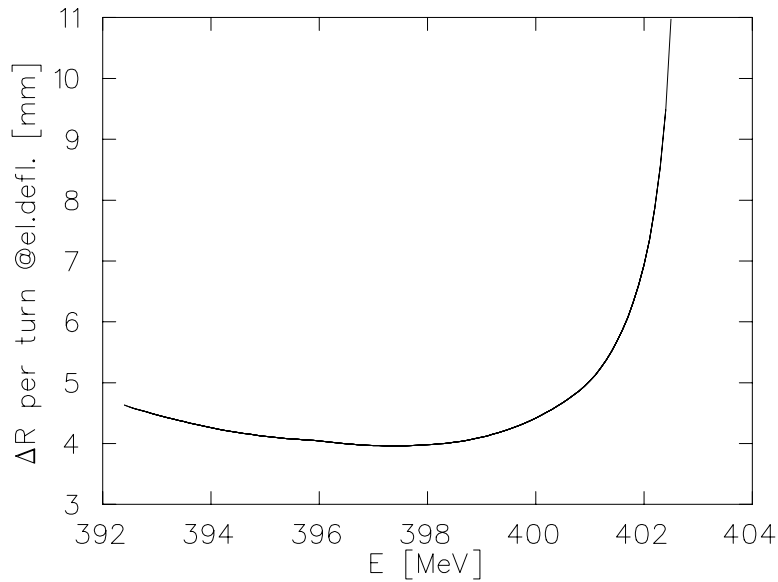

**Figure 24.** Radius gain per turn (in mm) due to acceleration (neglecting phase slip) as a function of the final energy (MeV/u). Observe the fast increase with energy near extraction.

### 3.3.2  Off-Centered Injection

If the beam is not injected on a centered trajectory, but rather on an off-centered one (as at PSI), the turn separation at extraction can be increased by selecting the appropriate phase and amplitude of the centering error. Figure 25 shows the Pr vs. R plot of the last three turns for three different ellipses in the radial plane. The solid curve represents a centered ellipse, the double-lined solid curve is a moderately off-centered (6 mm) ellipse, and the dashed ellipse is off-centered by 11 mm. The area of the ellipse corresponds to a normalized emittance of 0.4 π mm mrad. These three ellipses were tracked from the injection energy of 30 MeV/u to the final energy of ≈400 MeV/u. The phase of the centering error was adjusted to provide maximal turn separation at extraction.

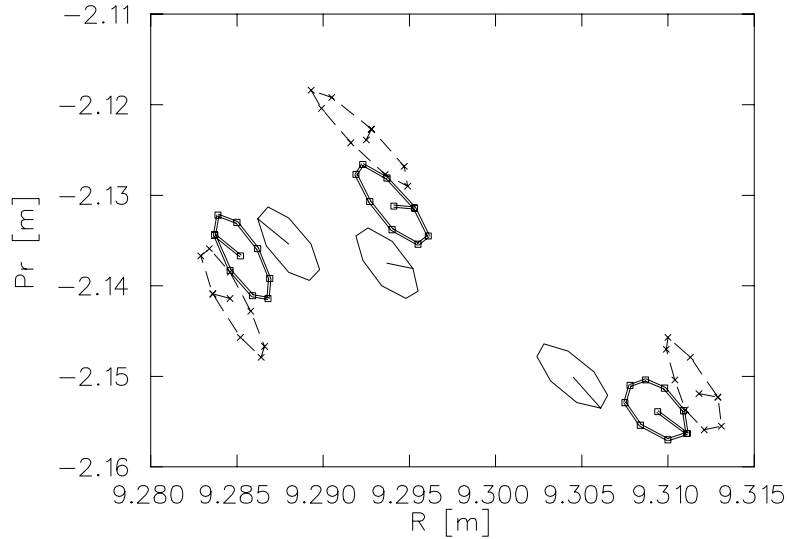

**Figure 25.** The last three turns before extraction at the electrostatic deflector entrance angle for three different off-centered beams with small (1mm, solid), moderate (6 mm, double solid) and large (11 mm, dashed) centering errors.

As seen in Figure 25, the ellipse for the large centering error is distorted, and hence, the moderate (6 mm) off-centering was selected for the rest of the calculations.

3.3.3 RF Voltage Variation

To illustrate the effect of the rf voltage variation, the ellipse with a moderate off-centering error from the previous analysis was accelerated with three different voltages; the nominal voltage and the nominal voltage ± 0.03 %. The value of this variation was chosen since it corresponds to the stability of the PSI rf system. Although small with respect to the turn separation, this effect will reduce the effective turn separation.

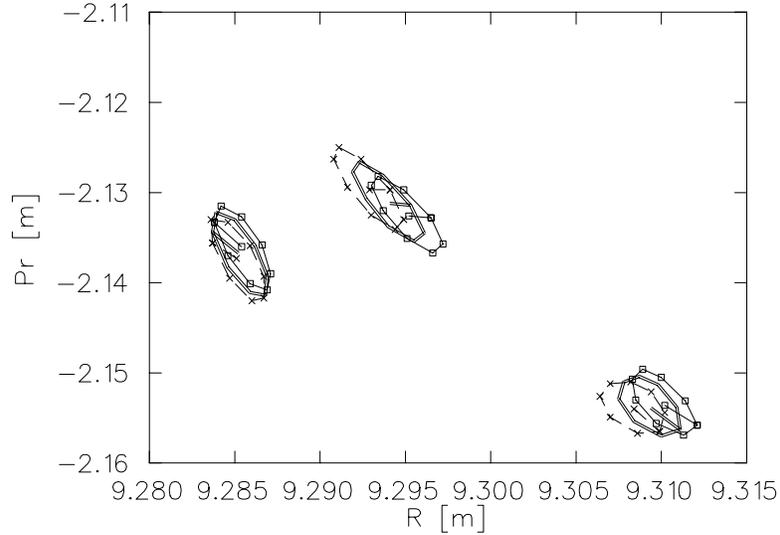

**Figure 26.** The last three turns before extraction at the deflector entrance for a beam that starts with moderate off-centering (6 mm) and with three different accelerating voltages given by the nominal voltage and the nominal voltage ±0.03%.

3.3.4   Bunch Length

The beam phase width was then incorporated into the simulations by tracking seven bunches of similar ellipses from the injection energy with a time variation of $5^o$ rf  ($\pm 15^o$ range). To include the effect of the injected beam energy spread, three groups with the nominal energy, and the nominal energy ± 0.1 % but dispersion matched were used. The total number of particles was 9 (populating each radial emittance ellipse) x 7 (different rf phases) x 3 (different energies) x 3 (different voltages) = 567. The results are show in Figure 27.

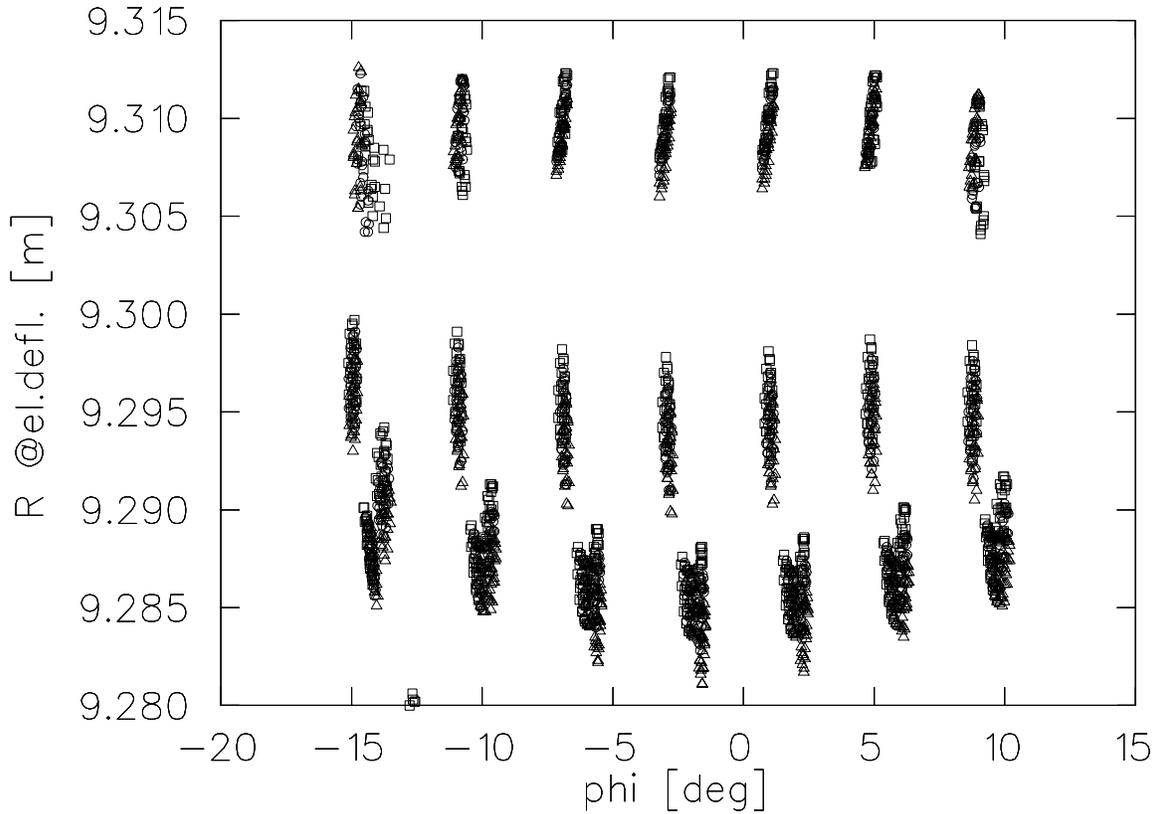

**Figure 27.** Radial separation between the extracted turn and the previous turns for an ensemble of 567 particles that include a phase space of 0.4 $\pi$ mm mrad normalized emittance, 7 phase groups that encompass ± 15° at injection, 3 different injection energies of nominal ±0.1%, and 3 different rf voltages (nominal±0.03%) . The particles are tracked from the injection energy to the entrance of the deflector where their position is plotted.

3.3.5   Transverse Emittance

From Figure 27, a turn separation of approximately 6 mm is achieved with most of the reduction in turn separation attributable to the extreme (± 15°) phase bunches. The same calculation was repeated with twice the radial phase space (0.8 $\pi$ mm mrad normalized); this result is given in Figure 28.

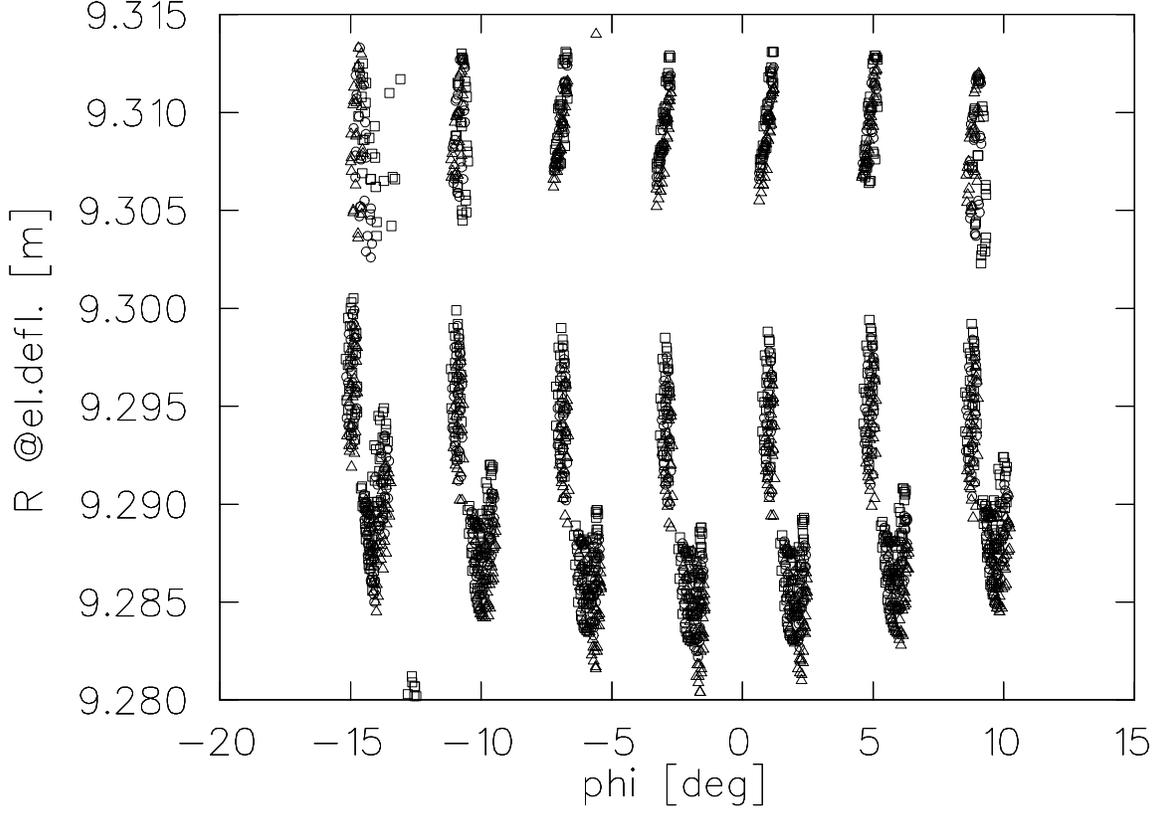

**Figure 28.** Similar to Figure 27 but with twice the nominal emittance in the radial phase space (0.8 π mm mrad).

From Figure 28, it was determined that after the inclusion of twice the transverse phase space, an rf voltage ripple, and an injection energy spread, there would be 5 mm separation between turns if the beam phase width is limited to ±10° in lieu of ±15°. This separation is more than adequate to provide the required high extraction efficiency.

3.3.6 Longitudinal Space Charge
The longitudinal space charge forces on the beam can also affect the beam size at extraction and hence, the extraction efficiency. Given the difficulty of accurately evaluating the effects of the longitudinal space charge force, two different approximations were employed to estimate the induced energy spread and consequently the effect on turn separation at extraction.

The first approach was to use a formula by Werner Joho[11] that provides an estimate of the total accelerating voltage spread caused by the longitudinal space charge force. In this model, the total space charge voltage spread ($V_{sc}$) is given by:

$$V_{SC} = 2800 <I> \frac{2\pi}{\Delta\phi} \frac{N^2}{\beta_f}$$

Where $<I>$ is the average current, N is the turn number, $\Delta\phi$ is the total phase width in rf degrees, and $\beta_f$ is the relativistic velocity at the final energy.

Using this model, the PSI Ring Cyclotron and the Separated Sector Cyclotron (SSC) parameters are compared in Table 7 where the value of Vsc is that predicted by the above formula and without any correction made by an rf frequency adjustment. Also given in Table 7, is the ratio of the voltage gain per turn ($V_{turn}$) from which the relative magnitude of the effect may be estimated. This ratio is significantly higher (2.3 vs. 0.32, i.e. 7 times worse) for the PSI cyclotron, making the PSI cyclotron worse relative to the effects of the longitudinal space charge force. These estimates provide confidence that the proposed Separated Sector Cyclotron (SSC) intensity will be compatible with separated turns.

| Value | PSI | SSC |
|---|---|---|
| $<I>$ (µA) | 1700 | 80 |
| N | 220 | 425 |
| $\Delta\phi$ (º) | 16 | 20 |
| $\beta_f$ | 0.79 | 0.715 |
| $V_{SC}$ (MV) | 6.56 | 1.0 |
| $V_{turn}$ (MV) | 2.9 | 3.2 |
| $V_{SC}/V_{turn}$ | 2.3 | 0.32 |

**Table 7.** Comparison of the longitudinal space charge force induced energy spread using the formalism of Joho[11] where smaller ratios of $V_{sc}/V_{turn}$ are better.

An alternative longitudinal space charge theory proposed by Gordon[12] was also evaluated. In this model, the beam's electric charge distribution is assumed to be distributed as a continuous sheet, and the electric field is calculated by assuming a trapezoidal charge distribution in azimuth. The electric field is integrated to calculate the total voltage spread. This calculation predicted an induced energy spread 14% larger than the Joho formula for the Separated Sector Cyclotron (SSC) and 4 % lower than Joho's for PSI. The agreement between the two approaches is quite good even though very different approximations where made in the two formalisms.

Using the results from Gordon's model, the frequency and initial phase were optimized to minimize the space charge energy spread. In the Separated Sector Cyclotron, the optimized space charge induced energy spreads were 0.12 and 0.13 MeV/u for beam heights of 1 cm and 0.5 cm respectively. These values are only ≈10% of the energy gain per turn, and therefore, would result in a relatively insignificant effect. For the case of the PSI Ring Cyclotron, the optimized energy spread was found to be equal to the energy gain per turn at a current of 0.73 mA. Since this is about one half the operating current for the achieved high-efficiency 1 MW extraction, it would suggest that these estimates are pessimistic. Further, even in the absence of this pessimism, the calculations would imply that the Separated Sector Cyclotron operating parameters will **not** present performance limitations due to space charge effects.

### 3.3.7 Horizontal-Vertical Coupling
The coupling of the horizontal and vertical motions can potentially decrease the turn separation at extraction by increasing the turn width through non-linear coupling. To address this issue, calculations were performed using the $Z^4$ code[13] that utilizes higher-order derivatives of the magnetic field. The simulation accelerates four particles on the periphery of a half ellipse (the motion is symmetric for the other half) in the z-pz plane and the maximum beam height as a function of energy is obtained.

Though the nominal normalized emittance from the linac is 0.4 $\pi$ mm mrad, to be conservative an emittance four times larger (1.6 $\pi$ mm mrad) was used in these simulations where the smaller and larger emittances would correspond to vertical beam sizes at injection of 0.5 and 1.0 cm respectively. From Figure 29, the maximum vertical beam size for each of the four particles is less than 2 cm near extraction. The increase in size is due to the decreasing vertical focusing frequency with increasing energy. The falling magnetic field near the extraction radius produces a rapid increase in the focusing frequency and hence, a beam size reduction near the extraction.

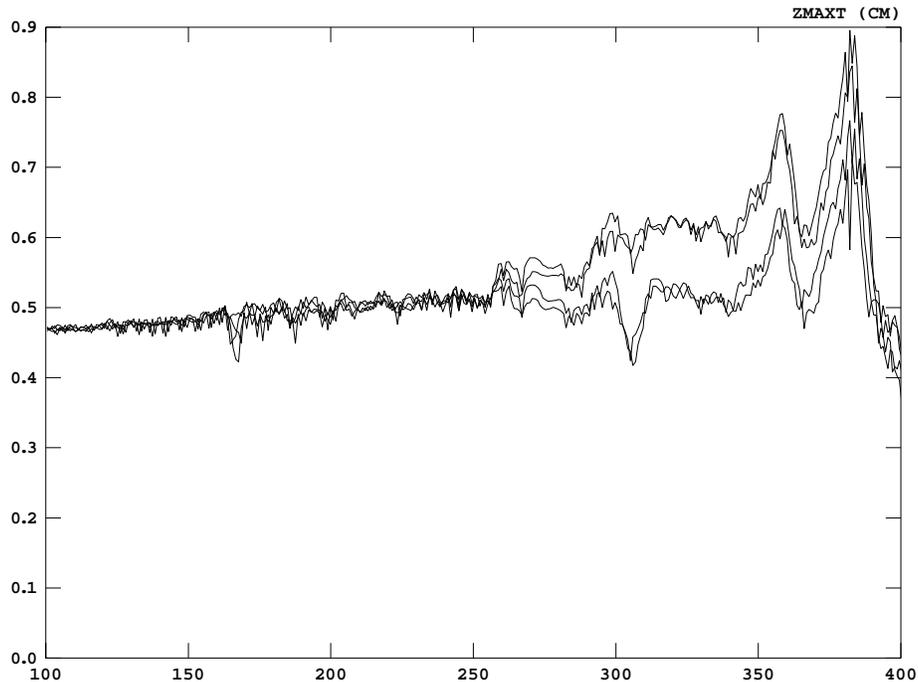

**Figure 29.** The maximum vertical beam size for each of four particles as function of energy for a normalized emittance of 1.6 $\pi$ mm mrad. The initial total beam height is ≈1 cm.

The coupling between the vertical motion of Figure 29 and that of the horizontal plane was determined by calculating the difference between the radial position of the four particles displaced from the median plane and a particle on the median plane (z = 0) all launched at the same horizontal position. The results of this analysis are given in Figure 30 from which it can be determined that near extraction the difference in radius is less than 0.5 mm; a negligible contribution to the turn broadening.

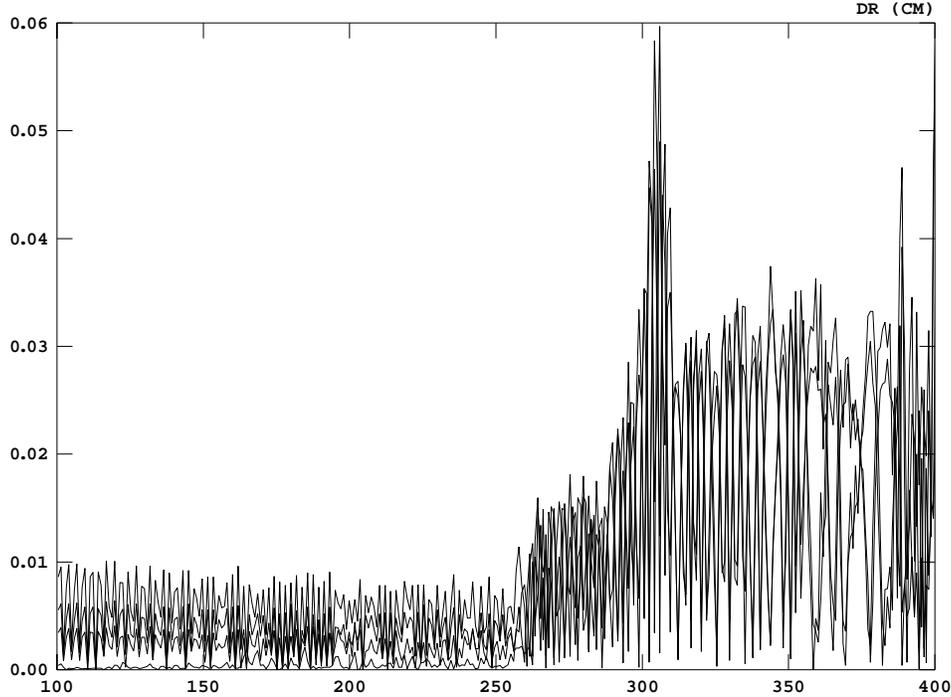

**Figure 30.** Radius difference between the particles with vertical (z) displacements and that of a particle that starts on the median plane. Four particles describing a half ellipse of appropriate for a normalized emittance of 1.6 $\pi$ mm mrad were tracked.

### 3.3.8 Extraction Elements

The extraction system will consist of the elements specified in Table 8. The SC (Superconducting) elements listed represent an inconsequential cryogenic load and should easily be accommodated by the cryogenic system required for the linac. The electrostatic septum will have a transverse thickness ($\approx$0.25 mm) small compared to the turn separation at extraction ($\approx$5 mm) and hence the extraction efficiency is predicted to be nearly 100%. The magnetic septum following has a septum width ($\approx$2.5 cm) again small compared to the turn separation ($\approx$5.5 cm) at that point. The remaining dipole and quadrupole elements that will complete the extraction process have standard parameters and will present no technical uncertainties.

| Element Type | Strength | Effective Length (m) |
|---|---|---|
| Electrostatic Septum | 60 kV/cm | 2 |
| Magnetic Septum | 2 T | 1.0 |
| SC Dipole | 3 T | 1.0 |
| Quadrupole | <35 T/m | 0.4 |
| SC Dipole | 3T | 1.0 |

**Table 8.** Extraction system elements.

# 4 Cost Estimates

The scope of the cost estimates developed below only encompasses the primary 400 MeV/u accelerator chain including the 30 MeV/u pre-accelerator and the Separated Sector Cyclotron. No attempt was made to include the cost of those elements reasonably common to both proposals (e.g. civil, utilities, etc.). The cost estimating process was done by dividing the overall project into two major areas; the 30 MeV/u pre-accelerator and the 400 MeV/u Separated Sector Cyclotron.

The costs for the pre-accelerator are derived from those presented by ANL at the first ISOL Technical Task Force.[2.] For the Separated Sector Cyclotron, a "bottom up" approach was used with the cost for each item developed in the following categories.

1. Non Recurring Costs (**NRC**) (e.g. engineering, setup, tooling, etc.)
2. Number Units Required
3. Price per Unit
4. Number of FTE-Years to Complete Task

The FTE costs were converted to dollars by assuming a rate of 150 K$ per FTE-Year.

## 4.1 Pre-Accelerator (PA)

| Item | Cost (M$) |
|---|---|
| RFQ | 1 |
| RT Linac | 1 |
| ≈250MV SC Linac | 43 |
| **SUBTOTAL** | **45** |

Table 9. Summary costs for the 30 MeV/u pre-accelerator system.

## 4.2 Separated Sector Cyclotron (SSC)

| Item | Cost (M$) |
|---|---|
| Injection System | 3.5 |
| Magnet System | 56.5 |
| RF System | 21.4 |
| Extraction System | 2.3 |
| Vacuum System | 3.6 |
| Diagnostic System | 2.3 |
| Specific Infrastructure | 5.8 |
| **SUBTOTAL** | **95.4** |

Table 10. Summary costs for a 400 MeV/u Separated Sector Cyclotron system.

As a cross check on the Separated Sector Cyclotron, the relative fractional costs of the main elements of the cyclotron system are compared in Table 11 to that realized for the PSI Ring and TRIUMF cyclotrons. This comparison shows that the cost distribution in the SSC is essentially the same as those of other extant large cyclotrons.

| Element | PSI (%) | TRIUMF (%) | MSU Separated Sector Cyclotron (%) |
|---|---|---|---|
| Magnet System | 56 | 49 | 59 |
| RF System | 17 | 17 | 22 |
| Other | 27 | 34 | 19 |

**Table 11.** Comparison of cost distribution for PSI, TRIUMF, and proposed Separated Sector Cyclotrons.

## 4.3 Totals

| Item | Cost (M$) |
|---|---|
| Pre-Accelerator System | 45 |
| Separated Sector Cyclotron | 95.4 |
| **TOTAL** | **140.4** |

**Table 12.** Total estimated cost for cyclotron based 400 MeV/u driver accelerator.

# 5 Summary

A design concept has been presented which will cost effectively meet the operational requirements for a 400 MeV/u acceleration system. Below are summary elements of the overall accelerator system.

**30 MeV/u Pre-Accelerator**
Approximately 250 MV of linac sections will provide a system suitable for producing 30 MeV/u heavy ions. The light ion variable energy criteria is met since this system will also produce protons energies of $\approx$145 MeV and $^4$He energies of $\approx$90 MeV/u (360 MeV total). Three stripper locations will be required to strip all ions to a Q/A $\approx$ 1/3 though each beam will be stripped only once. The longitudinal dynamics are well matched to the Separated Sector Cyclotron requirements.

**Separated Sector Cyclotron**
A high-efficiency system to inject beam from the linac into the Separated Sector Cyclotron has been described.

It was determined that a Q/A range of Q/A = 1/3 ± 10.1% would provide no limitations on the fragment intensities and that this range was achievable for the Separated Sector Cyclotron.

A vacuum of ≈$10^{-7}$ torr was found to provide adequate transmission with the power loss estimated to be only ≈170 W under this condition.

An rf system largely based on the PSI design was determined to meet design requirements.

The extraction system efficiency was simulated including the effects of larger than nominal transverse emittance, rf voltage ripple, and injected beam energy spread. With an rf phase width of ±10° the clear space between turns at extraction is 5 mm. Additional effects caused by longitudinal space charge forces and horizontal-vertical motion coupling were also evaluated and found to be acceptably small effects. Hence, a high-efficiency extraction system in the range of 99.99% is achievable for the Separated Sector Cyclotron design presented.

A "bottom up" costing of the Separated Sector Cyclotron system was done and the total estimated cost for this element of the accelerator chain was found to be <100 M$ and the full acceleration chain to be <150 M$.

# 6   References

[1] F. Marti, et al., "A Cyclotron Based Acceleration Chain for a 400 MeV/u Driver for a Radioactive Beam Facility", MSUCL-1126, June 1999.

[2] ISOL Technical Task Force (June 24-25, 1999 at NSCL), C. Leemann, Chairman.

[3] E. Baron, M. Bajard and Ch. Ricaud, "Charge exchange of very heavy ions in carbon foils and in the residual gas of GANIL cyclotrons", NIM, **A328** (1993), 177-182.

[4] Private communication with Peter Ostroumov.

[5] R. Servranckx, "Users Guide to the Program DIMAD", SLAC REPORT 285, UC-28(A), May 1985.

[6] B. F. Milton, "CYCLONE VER8.3", TRI-DN-99-4, January 28, 1999.

[7] B. Franzke, Vacuum Requirements for Heavy Ion Synchrotrons, IEEE Trans. on Nuc. Sci. **NS-28** No.3, (1981) 2116.

[8] N. Bohr and J. Lindhard, Dan. Mat. Fys. Medd. **28**, no. 7 (1954).

[9] H.D. Betz and Ch. Schmelzer, UNILAC 1-67 (1967).

[10] H.D. Betz, G. Hortig, E. Leischner, Ch. Schmelzer, B. Stadler, and J. Weihrauch, Phys. Lett. **22** No. 5 (1966) 643

[11] W. Joho, "High Intensity Problems in Cyclotrons", 9$^{th}$ Int. Conference on Cyclotrons, p337, Caen, France 1981.

[12] M.M. Gordon, "The longitudinal space charge effect and energy resolution", 5$^{th}$ Intl. Cyclotron Conf., pages 305-317, Butterworths, London, 1971.

[13] M.M. Gordon and V. Taivassalo, "The $z^4$ code and the focusing bar fields used in the extraction calculations for superconducting cyclotrons", Nucl. Inst. and Meth., **A247**, 423 (1986).